\documentclass{jpp}

\usepackage[T1]{fontenc}
\usepackage[utf8]{inputenc}
\usepackage{amsmath}
\usepackage{subcaption}
\usepackage{amssymb}
\usepackage{graphicx}
\usepackage{esint}
\usepackage{float}
\usepackage{tikz}

\usepackage{mathtools}
\usepackage{physics}
\usepackage{hyperref}
\usepackage[dvipsnames]{xcolor}
\usepackage[english]{babel}
\usepackage[autostyle]{csquotes}
\usepackage{placeins}

\usepackage{setspace}
\usepackage{titlesec}
\usepackage{booktabs}

\usepackage{multirow}
\usepackage{threeparttable}
\usepackage{threeparttablex}
\usepackage{tabularx}
\usepackage{tabulary}

\usepackage{upgreek}

\newcommand \s {s}
\newcommand \maxwellians {F_{0s}}
\newcommand \maxwellian {F_{0}}
\newcommand{\rmd}{\mathrm{d}}
\renewcommand{\vec}{\boldsymbol}

\newcommand {\df}{\delta \! f}
\newcommand {\dg}{h_{s}}
\newcommand {\dgg}{h}

\newcommand \besselks {J_{0 \s, \boldsymbol{k}}}
\newcommand \besselk {J_{0, \boldsymbol{k}}}

\newcommand \vpa {v_{\parallel}}
\newcommand {\bk}{k}
\newcommand {\x}{\psi}
\newcommand {\y}{\alpha}
\newcommand {\kx}{k_{\x}}
\newcommand {\ky}{k_{\y}}
\newcommand {\bky}{\bk_{\y}}
\newcommand {\bkx}{\bk_{\x}}

\newcommand \gyrophi {\langle \phi \rangle_{\vec{R}}}

\newcommand \xperp {\hat{v}_{\perp}^2}

\newcommand \omegap {\omega^{\prime}}

\newcommand \omegadxy {\omega_{d, \bkx \bky}}
\newcommand \omegadx {\omega_{d, \bkx}}
\newcommand \omegady {\omega_{d, \bky}}
\newcommand \omegadmy {\omega_{d, -\bky}}

\newcommand \hatomegadxy {\hat{\omega}_{d, \bkx \bky}}
\newcommand \hatomegadx {\hat{\omega}_{d, \bkx}}

\newcommand \hatomegadmy {\hat{\omega}_{d, -\bky}}

\newcommand \omegastary {\omega_{*, \bky}}
\newcommand \omegastarmy {\omega_{*, -\bky}}

\newcommand \hatomegastary {\hat{\omega}_{*, \bky}}
\newcommand \hatomegastarmy {\hat{\omega}_{*, - \bky}}

\newcommand \veck {\vec{k}}
\newcommand \hq {\hat{h}}

\newcommand \vths {v_{\text{th}, \s }}

\newcommand \bbk {b_{\veck}}

\newcommand \veckp {\veck^{\prime}}

\newcommand \besseltrio {J_{0, \bkx} J_{0, \bky}J_{0, \bkx\bky}}

\hypersetup{
  colorlinks,
  linkcolor={blue!80!black},
  citecolor={blue!80!black},
  urlcolor={blue!80!black},
}

\title{A Weakly Nonlinear Theory of Zonal-Flow Forcing in Gyrokinetic Turbulence}
\author{ Georgia Acton, Eduardo Rodr\'iguez, Gareth Roberg-Clark, Alessandro Zocco}

\affiliation{Max-Planck-Institut f\"ur Plasmaphysik, Wendelsteinstra{\ss}e
1, D-17491 Greifswald, Germany}

\begin{document}
\maketitle

\begin{abstract}
The forced generation of zonal flows by microinstability-driven turbulence is investigated within the framework of local gyrokinetic theory far from marginality. We use a numerically and physically informed three-wave truncation scheme, which allows the prediction of the zonal-flow $\kx$-spectrum during the early phase of nonlinear gyrokinetic simulations. 
The model reproduces the known 2$\gamma$ growth rate resulting from nonlinear beating of linearly unstable primary modes, in line with previous results, without any marginal stability point.
The phase-space structure of such zonal flow is strongly constrained by that of the driving fluctuations, which is essential to understand its behaviour in the region of validity. It is shown that this leads to an enhanced residual spectrum compared to the classic Rosenbluth–Hinton calculation.
\end{abstract}
\section{Introduction}

Zonal flows (ZFs) have become a major subject of investigation in the field of magnetically confined fusion plasmas (see e.g., \cite{Hasegawawakatani,RosenbluthHinton1998, Diamond_2005, ItohItohDiamondetal}). 
They are defined as self-organised, turbulence-driven flows that are largely constant over flux surfaces, while remaining radially sheared. For a magnetic field of the form $\mathbf B=\nabla\alpha\times\nabla\psi$ \citep{kruskalkulsrud}, where $\alpha$ labels magnetic field lines and $\psi$ flux surfaces, this behaviour can be described by wave-vectors $k_{\alpha} \equiv 0$, and $k_{\psi}$ on gyrokinetic scales. The combination of large-scale coherence and radial differential flow enables ZFs to suppress turbulence through radial decorrelation, both in axisymmetric and non-axisymmetric magnetic configurations \citep{biglari1990influence, BillGreg1993, lin1998turbulent, terry2000suppression, tiwari2025zonal}. Driven by this promise, the highly nontrivial character of ZFs has fueled the imagination of many and generated a considerable amount of literature (see e.g., \cite{Diamond_2005,ItohItohDiamondetal}). 

The seminal work of Rosenbluth and Hinton \citep{RosenbluthHinton1998} remains a first simple reference point for understanding ZFs. It uses linear, radially local, electrostatic gyrokinetic theory \citep{friemanchen} to analytically derive how much of an initial ZF remains at late times. What remains is the long-wavelength, non-Landau damped component, commonly referred to as the \textit{ZF residual}. This remanent level is often used to characterise ZFs (and the quality of the magnetic field which supports it), partly because it measures how \enquote{resilient} the ZF is, and partly because it is one of the few analytical results available (see e.g., \cite{watanabe2008regulation, mishchenko2008collisionless, xanthopoulos2011zonal, helander2011oscillations,  monreal2016residual, plunk2024residual, goodman2024quasi}). It has been suggested that the long time residual, $\phi_\infty$, may facilitate larger ZFs in nonlinear scenarios (see e.g., \cite{waltz2008numerical, mynick2010optimizing, st-onge_2017, plunk2024residual}), however, a definitive link between the residual value and the prevalence of ZFs nonlinearly has yet to be firmly established.

However, one may ponder about the initial generation of ZFs. Two main generation mechanisms have been identified: generation through secondary (or modulational \citep{strintzi2007relation}) instabilities (see e.g., \cite{Rogers2000, smolyakov2000zonal, plunk2017nonlinear, Richard2024}), and generation via direct nonlinear forcing. Much of the modern focus has been on the former, which describes how perturbations to non-zonal ion-scale turbulence nonlinearly generate shear flows \citep{Rogers2000, ChenLinWhite}. It is often appropriate to model the underlying non-zonal turbulence as streamers \citep{Steve1991}; \textit{i.e.}, self-organised, radially-elongated ($k_{\psi}=0$) structures. This scenario was first formalised by Rogers–Dorland–Kotschenreuther (RDK) \citep{Rogers2000} in a gyro-fluid, slab model. They showed that the ZF grows super-exponentially, akin to a Kelvin-Helmholtz instability. Recently this work has been extended by \cite{Richard2024} to include toroidal effects using a gyrokinetic framework. They emphasise the importance of the radial component of the magnetic drift, which leads to an additional zonal growth path through the toroidal secondary mode (TSM). Both of these descriptions pertain to a strongly nonlinear regime of the turbulence evolution, where only the initial streamer behaves linearly. Other treatments of this nonlinear generation mechanism do not necessarily consider the streamer assumption, and use a finite drift wave frequency for it \citep{ChenLinWhite}.

The second mechanism of ZF generation is the more direct, purely driven scenario. In this direct drive the growing non-zonal modes beat together to nonlinearly drive the ZF. This has been repeatedly observed by various authors as a \enquote{$2\gamma$ growth}, but it is often considered a lesser effect \citep{Rogers2000, qiu2016effects}. Although initial attempts have been made to combine these two mechanisms \citep{ChenQiuZonca2024}, in this paper we present a careful analysis of the purely driven mechanism\footnote{We make a comparison between work presented in this model, and this existing work is presented in Appendix~\ref{sec:comparison_to_previous_wrk}.}.

In Section~\ref{sec:Framework} we introduce the gyrokinetic framework, on which our work is built. Following this, in Section~\ref{sec:Regimes}, we identify a number of regimes based off systematic orderings of amplitudes and time scales, and construct a truncated minimal-mode model for each. We find the secondary mechanism manifests in a \enquote{strongly nonlinear} regime in line with \cite{Rogers2000} and \cite{Richard2024}. Whereas, the forced mechanism dominates in what we call a \enquote{weakly nonlinear} regime. Thereafter we focus on the latter, whose suitability we numerically validate in Section~\ref{sec:weaklyNL}, and analytically study in Section~\ref{sec:Laplace}. The known $2\gamma$ growth rate is shown to be the dominant response, with no marginal points. Following this we analytically derive its amplitude, and further investigate its structure in Section~\ref{sec:res_2_gamma} by studying its residual spectrum. We show that its velocity-space structure significantly impacts the residual level relative to the classic \cite{RosenbluthHinton1998} result. We close with some conclusions.

\section{Gyrokinetic Framework}\label{sec:Framework}

We consider turbulence associated with fluctuations of the electrostatic potential $\phi(\vec{r},t)$ and the distribution function $\delta\!f_{\s}(\vec{r},\mu,\mathcal{E},t)$. Here, we have introduced the phase-space coordinates $\{t,\mathcal{E},\mu,\vartheta,\vec{r}\}$, corresponding to time, energy per mass $\mathcal{E}=v^2/2$, the first adiabatic invariant $\mu=v_\perp^2/(2B)$, gyrophase, and particle position, respectively. The full distribution function is decomposed as
\begin{equation}
f_{\s} = \maxwellians + \delta \! f_{\s}
\equiv \maxwellians\!\left(1 - \frac{q_{\s}\phi(\vec{r},t)}{T_{0s}}\right)
+ \dg(\vec{R}_{\s},\mu,\mathcal{E},t)
+ \mathcal{O}(\rho_*^2),
\end{equation}
for species $\s,$ with charge $q_{\s},$ and temperature $T_{0s}$. Here, $\maxwellians$ is the equilibrium distribution function, and is taken to be a Maxwellian with total density $n_{s}=\int d^{3}\mathbf{v}f_{s}$. We also define $\rho_* = \rho_i/L \ll 1$ to be the ratio of the thermal ion gyroradius to the system size. The above satisfies the following gyrokinetic (GK) ordering
\begin{equation}
\frac{\delta \! f_{\s}}{\maxwellians} \sim \frac{q_{\s} \phi}{T_{0, \s}} \sim \frac{k_\parallel}{k_\perp}
\sim \frac{\omega}{\Omega_{\s}} \sim \rho_* \ll 1, \label{eq:GK_ordering}
\end{equation}
where $k_\parallel$ and $k_\perp$ are wavevectors parallel and perpendicular to the equilibrium magnetic field. We also introduce a characteristic frequency of fluctuations, $\omega$, and the gyrofrequency, $\Omega_{\s} = q_{\s} B/ m_{\s} c$, with $B$ the magnetic field magnitude, and $m_{\s}$ the species mass. Here, $\dg$ denotes the non-adiabatic part of the distribution function, evaluated at the gyro-centre position $\vec{R}_{\s}$. The gyro-centre and particle positions are related by $\vec{R}_{\s} = \vec{r} - \boldsymbol{\rho}_{\s}(\vartheta)$, where $\boldsymbol{\rho}_{\s}(\vartheta)
= \hat{\boldsymbol{b}} \times \boldsymbol{v} / \Omega_{\s}$
is the velocity-dependent gyroradius, and $\hat{\boldsymbol{b}} \equiv \vec{B}/|B|$ is the unit vector in the direction of the magnetic field. Note that the electrostatic potential is evaluated at the particle position, while $\dg$ is evaluated at the gyro-centre. Together, the potential and distribution function satisfy the quasineutrality condition \citep{friemanchen, brizard2007foundations, abel2013multiscale}
\begin{equation}
    \sum_{s} q_{\s} \int \rmd^3\mathbf{v} \left[ \langle \dg \rangle_{\vec{r}} - \frac{q_{\s}}{T_{0s}} \maxwellians \phi \right] = 0,
    \label{eq:QN_real}
\end{equation}
where the velocity integral is evaluated at constant particle position, $\vec{r}$. The angle brackets denote a gyroaverage, which we define for a generic function, $g$, as
\begin{subequations}
\begin{align}
	\langle g(\vec{r}) \rangle_{\vec{R}_{\s}}
	&= \left\langle g(\vec{R}_{\s} + \boldsymbol{\rho}_{\s}(\vartheta)) \right\rangle_{\vec{R}_{\s}}
	= \frac{1}{2\pi}\int_0^{2\pi}
	g\!\left(\vec{R}_{\s} + \boldsymbol{\rho}_{\s}(\vartheta)\right)\,\rmd\vartheta,
	\\
	\langle g(\vec{R}_{\s}) \rangle_{\vec{r}}
	&= \left\langle g(\vec{r} - \boldsymbol{\rho}_{\s}(\vartheta)) \right\rangle_{\vec{r}}
	= \frac{1}{2\pi}\int_0^{2\pi}
	g\!\left(\vec{r} - \boldsymbol{\rho}_{\s}(\vartheta)\right)\,\rmd\vartheta,
	\label{eq:gyroaverage_particle_position_constant}
\end{align}
\end{subequations}
where $\vec{R}_{\s}$ and $\vec{r}$ are held fixed, respectively, during the
$\vartheta$ integration. The non-adiabatic distribution function $\dg$ satisfies the nonlinear
GK equation \citep{friemanchen}
\begin{multline}
    \left( \frac{\partial}{\partial t}
    + v_\parallel \nabla_\parallel
    + \mathbf{v}_{d,s}\cdot\nabla
    \right)\dg =
    \frac{q_{\s} \maxwellians}{T_{0s}}
    \frac{\partial}{\partial t}
    \left\langle \phi \right\rangle_{\vec{R}_{\s}}
    \\
    -\frac{c}{B}\hat{\mathbf{b}}\cdot
    \nabla\left\langle \phi \right\rangle_{\vec{R}_{\s}}
    \times\nabla \maxwellians
    -\frac{c}{B}\hat{\mathbf{b}}\cdot
    \nabla\left\langle \phi \right\rangle_{\vec{R}_{\s}}
    \times\nabla\dg ,
    \label{eq:GK_real}
\end{multline}
where $\grad_{\parallel} = \hat{\vec{b}} \cdot \grad$ indicates a derivative along fieldline, and $\mathbf{v}_{d,\s}$ the magnetic drift. 

Leveraging the ordering in equation~\eqref{eq:GK_ordering}, a WKB representation is used to treat the perpendicular directions in equations~\eqref{eq:QN_real} and \eqref{eq:GK_real}. We define the perpendicular wave-vector $\vec{k}_\perp=k_\alpha\nabla\alpha+k_\psi\nabla \psi$, with the local perpendicular wavenumbers $\{k_\psi, k_\alpha\}$ constant. This yields the working equations for quasineutrality 
\begin{equation}
	\sum_{s}\frac{1}{n_{0s}}\int d^{3}\mathbf{v}\,J_{0s,\vec{k}}h_{s,\vec{k}}=\sum_{s}\frac{q_{s}\phi_{\vec{k}}}{T_{0s}}\,,
\end{equation}
and the GK equation
\begin{equation}
    \mathcal{L}_{\vec{k}} \, \dgg_{\s, \vec{k}} = \mathcal{R}_{\vec{k}} \frac{q_{\s} \maxwellians}{T_{0\s}} J_{0s,\vec{k}} \phi_{\vec{k}} - \mathcal{N}_{\vec{k}},
    \label{eq:schematic_GK}
\end{equation}
where
\begin{subequations}
\begin{align}
        \mathcal{L}_{\vec{k}} &= i\partial_t + iv_\parallel \nabla_\parallel -\vec{k}_\perp\cdot\mathbf{v}_{d, \s}
        \\
        \mathcal{R}_{\vec{k}} &= i\partial_t - c \frac{k_\alpha T_{0s}}{q_{\s}} \frac{\maxwellians'}{\maxwellians} 
        \\
        \mathcal{N}_{\vec{k}} &= ic\sum_{\vec{k}'} \left(k_\psi k_\alpha' - k_\alpha k_\psi'\right) \besselks \phi_{\vec{k}-\vec{k}'} \dgg_{s, \vec{k}'}
\end{align}
\end{subequations}
where the gyroaverages have given Bessel
functions \citep[Chapter 9]{abramowitz1948handbook}, $\besselks=J_{0}(\sqrt{2b_{\vec{k}}} v_{\perp}/v_{\mathrm{th, \s}}),$ with $b_{\vec{k}} = \vec{k}^2 \rho_i^2/2$, where we have dropped the perpendicular subscript on $\vec{k}$ for ease of notation. We do likewise with the species subscript from here on, and consider electrons to adopt a modified Boltzmann response (see e.g., \cite{BillGreg1993} Eq.~(5); \cite{hammett1993developments, abel2013multiscale}). For this the electron distribution function is taken as $\delta f_e = e (\phi-{\langle \phi \rangle}_{\x})/T_e$, where $\langle \cdot \rangle_{\psi} \equiv 1/V \int dz \int d\alpha \cdot / (\hat{\bf{b}} \cdot \nabla z)$ indicates a flux-surface average, with $V$ the volume of the flux-surface.
\section{Regimes}
\label{sec:Regimes}

\newsavebox{\leftbox}
\begin{figure}
    \centering
    \begin{minipage}[t]{0.345\textwidth}
        \centering
        \vspace{0.075cm}    
        \includegraphics[width=\linewidth,keepaspectratio]{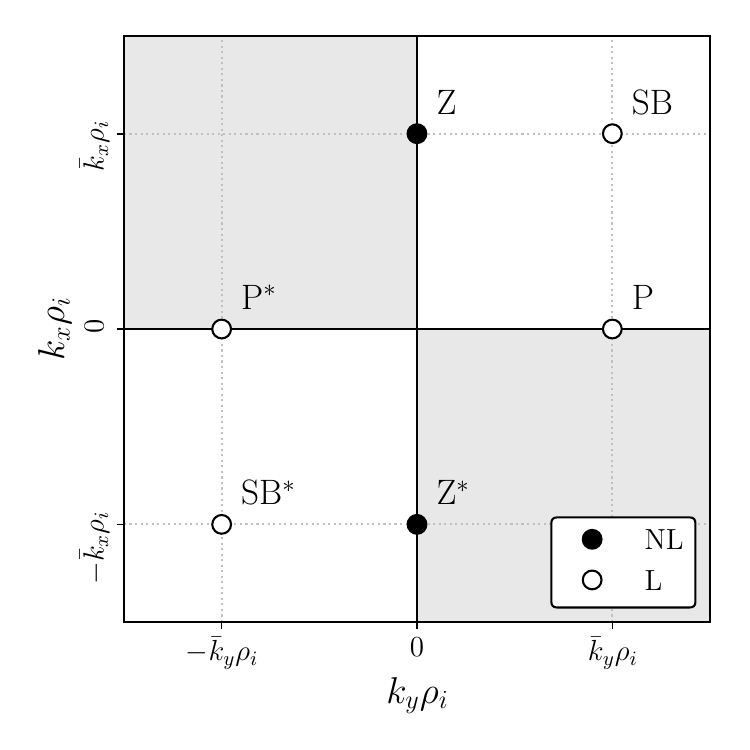}
    \end{minipage}
    \begin{minipage}[t]{0.64\textwidth}
        \vspace{0pt}
        \centering
        \includegraphics[width=\linewidth,height=0.40\textheight,keepaspectratio]{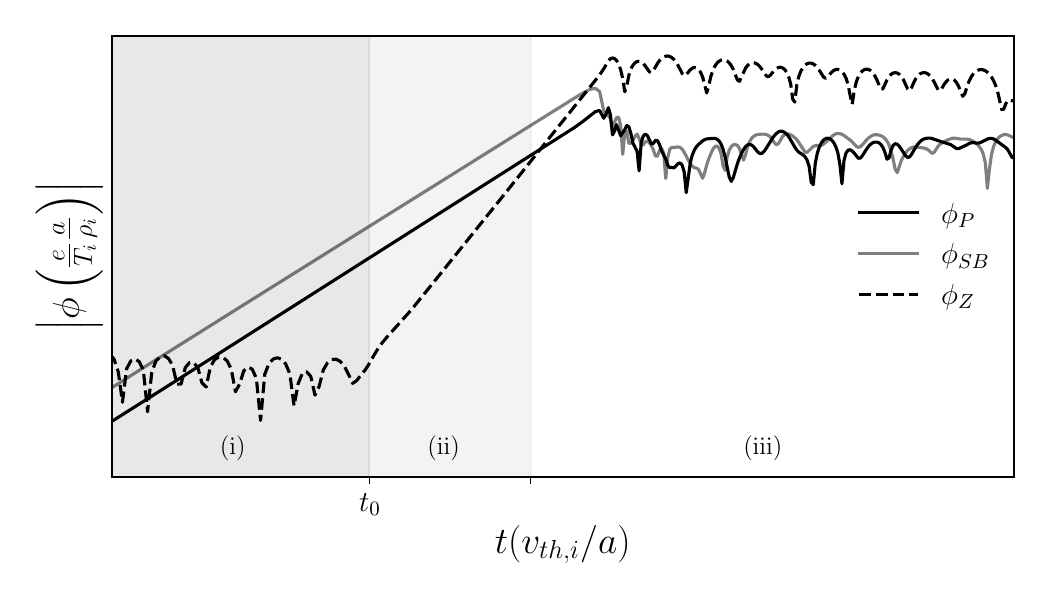}
    \end{minipage}

    \caption{\textbf{Minimal Model and Regimes}. (a) $\vec{k}$-space diagram illustrating the absolute minimum number of modes required to capture the nonlinear dynamics of the ZF. This is the model used to describe the weakly nonlinear regime.
    (b) Illustrative time trace of the evolution of the three basic modes (see (a)) in a nonlinear GK simulation. Intervals over which different dynamical regimes are valid are labelled; region (i) corresponds to the linear regime, region (ii) to the weakly nonlinear regime, and region (iii) encompasses both the strongly nonlinear and fully nonlinear regimes. }
  \label{fig:compare_regimes}
\end{figure}

The main difficulty of the GK equation, Eq.~\eqref{eq:schematic_GK}, is the presence of the nonlinear term $\mathcal{N}_{\vec{k}}$, related to the advection of perturbations by the perturbed electrostatic potential. In the modal language of this paper, this nonlinear term links the evolution of a mode $\vec{k}$ to all other pairs satisfying the convolution condition, $\vec{k}_1+\vec{k}_2=\vec{k},$  for two given wavenumbers $\vec{k}_1, \vec{k}_2$ excluding self-interactions. This complexity prevents one from explicitly solving such an equation in its full form. 

To make progress towards understanding evolution and behaviour of the zonal mode, we start by considering a reduced number of modes. Such minimal-models are capable of reproducing key numerical results with good accuracy in specific regimes, as we will demonstrate later and has been seen by others (see e.g., \cite{guillon2025phase, Rogers2000}).
The absolute miminimal number necessary for capturing the nonlinear dynamics is clearly three and their complex conjugates; zonal ($k_{\psi}\neq0,\,k_{\alpha}=0$), primary $(k_{\psi}=0,\,k_{\alpha}\neq0)$, and sideband $(k_{\psi}\neq0,\,k_{\alpha}\neq0)$ (see Figure~\ref{fig:compare_regimes}). Their associated electrostatic potentials are denoted by $\phi_{\kx}$, $\phi_{\ky}$, and $\phi_{\kx \ky}$ respectively. The minimal form of the truncated problem is three equations of the form of Eq.~\eqref{eq:schematic_GK} (one for each mode), plus quasineutrality for each. 

The dynamics of the ZF are determined by the structure of these equations, which in turn depends strongly on the relative importance of the linear and nonlinear terms. Different relative orderings imply distinct dominant dynamics, and can therefore be described as different \textit{regimes} of ZF evolution. The most basic is a fully \textit{linear regime}: the nonlinear terms in all equations play a subdominant role. This regime is illustrated as region (i) in Figure~\ref{fig:compare_regimes}: the initial phase in a typical evolution of a nonlinear GK simulation. When the zonal mode evolves nonlinearly, but the other two modes evolve linearly, then a \textit{weakly nonlinear regime} ensues. The occurrance of such a regime is illustrated as region (ii) in Figure~\ref{fig:compare_regimes}. Finally, \textit{strongly nonlinear} and \textit{fully nonlinear regime}s are here defined as a system for which two or all three mode types are strongly influenced by their nonlinear terms respectively. The latter is illustrated in region (iii) in Figure~\ref{fig:compare_regimes}. We now discuss these regimes as they pertain to zonal flows.

\subsection{Linear Regime}

In the linear regime, every mode is taken to evolve linearly. The linear evolution of the zonal mode is a well studied problem, corresponding to setting $\mathcal{N}_{\kx} = 0$ in Eq.~\eqref{eq:schematic_GK}. It is within this regime that the classic Rosenbluth-Hinton residual is derived \citep{RosenbluthHinton1998}. 
    
To argue when this scenario is dominant we need to consider the ordering of different terms in the nonlinear GK equation. The linear term is ordered as $\mathcal{L}_{k}\sim\omega_k^\mathrm{lin}$, where $\omega_k^\mathrm{lin}$ represents a linear time scale. We do not distinguish at this level between the different terms that make up the linear operator. With that, a linearly evolving zonal flow requires
\begin{equation}
    \omega_{\kx}^\mathrm{lin}h_{\kx}\gg\mathcal{N}_{\kx}.
\end{equation}
To make the ordering of the different terms more explicit, let us introduce additional simplifications and define adimensional quantities. It is convenient in this context to take $h \sim (q\phi/T)F_0$, such that the above ordering can be written as,
\begin{equation}
    \tilde{\phi}_{\kx}\gg  \frac{\tilde{\mathcal{N}}}{\tilde{\omega}_{\kx}^\mathrm{lin}} \tilde{\phi}_{\ky}\tilde{\phi}_{\kx\ky}, \label{eq:ord_linear}
\end{equation}
where $\tilde{\phi}\sim \rho^{-1}_* (q\phi/T)$, $\tilde{\omega}=\omega a/\vths$ and $\tilde{\mathcal{N}}\sim B_0 \kx\rho_i\ky\rho_i$, with $B_0$ a reference magnetic field. For the linear regime to hold, we thus need the amplitude of the zonal mode to be sufficiently large compared to the product of the primary and sideband.

However, the inequality given in Eq.~\eqref{eq:ord_linear} is bound to be violated given linearly unstable non-zonal modes. Linearly, the zonal modes cannot grow. In the absence of parallel streaming, under the local approximation, the linear modes are all damped on a time scale $\sim \omega_{d, \kx}$. In the nonlocal limit, if one retains the effects of parallel streaming, one expects the system to damp geodesic acoustic modes (GAMs) \citep{Winsor1968,Qiu2019} via Landau damping. In fact, all components of the ZF decay except the residual \citep{RosenbluthHinton1998}. Meanwhile, finite $\ky$-modes can grow linearly, as they can tap into equilibrium gradients as sources of free energy (\textit{e.g.}, the ion temperature gradient driven instability, ITG). 
Hence, even if initially the system is fully linear, at some time $t_0$ the non-zonal modes will eventually become sufficiently large to drive the nonlinear term. 

\subsection{Weakly Nonlinear Regime}\label{sec:intro_weak}

Following the linear discussion above, it is natural to consider the weak regime as one in which the zonal mode comes to being nonlinearly dominated, while the sideband and primary remain linear. This corresponds to region (ii) in Figure \ref{fig:compare_regimes}. 

To model this regime we adopt the minimal three-mode (triadic) model, see Figure~\ref{fig:compare_regimes}a, with the following simplified nonlinearities in the GK equations, Eq.~\eqref{eq:schematic_GK}
\begin{align}
	\begin{cases}
        \displaystyle
		\mathcal{N}_{\ky} & = 0, \\
        \displaystyle
		\mathcal{N}_{\kx \ky} & = 0, \\
		\displaystyle
		\mathcal{N}_{\kx} & = ic \kx \ky \left[ J_{0,\ky}\,\phi_{-\ky}\,\dgg_{\kx \ky} - J_{0,\kx \ky}\,\phi_{\kx \ky}\,\dgg_{-\ky}
		\right].
	\end{cases}
\end{align}
To understand when this regime is valid, we must once again consider the orders of the various terms in the GK equation. The non-zonal modes evolve linearly, and for simplicity we take them to do so on the same time scale, $\omega_{\kx\ky}^\mathrm{lin}\sim\omega_{\ky}^\mathrm{lin}$, although this assumption is not essential. By contrast, the characteristic time scale of the ZF is set by the nonlinear drive, which imposes its own timescale. Since the nonlinear drive is generated by modes each growing like $\phi_{\ky} \sim \exp(i \gamma_{\ky}^\mathrm{lin})$, the nonlinear forcing goes like $\phi_{\ky} \phi_{\kx, \ky}\sim \exp(2 i \gamma_{\ky}^\mathrm{lin})$, meaning that the ZF evolution time scale is ordered $\omega_{\kx}^\mathrm{NL}\sim \gamma_{\ky}^\mathrm{lin}$. 

For the nonlinear term to dominate we need,
\begin{equation}
    \tilde{\phi}_{\kx}\ll \frac{\tilde{\mathcal{N}}}{\tilde{\omega}_{\kx}^\mathrm{lin}} \tilde{\phi}_{\ky}\tilde{\phi}_{\kx\ky}, 
    \label{eq:weakly_nl_zonal_ineq}
\end{equation}
in which case we can order $\tilde{\phi}_{\kx}\sim \tilde{\mathcal{N}} \tilde{\phi}_{\ky}\tilde{\phi}_{\kx\ky}/\gamma_{\ky}^\mathrm{lin}$. We still need the nonlinear terms in the primary and sideband equations to remain small, and thus insist the following, (invoking $\omega_{\kx\ky}^\mathrm{lin}\sim\omega_{\ky}^\mathrm{lin}\sim \gamma_{\ky}^\mathrm{lin}$ directly)
\begin{subequations}
\begin{align}
    \tilde{\phi}_{\ky}\gg & \frac{\tilde{\mathcal{N}}}{\tilde{\omega}_{\ky}^\mathrm{lin}} \tilde{\phi}_{\kx\ky}\tilde{\phi}_{\kx}\sim \left(\frac{\tilde{\mathcal{N}}}{\tilde{\omega}_{\ky}^\mathrm{lin}}\right)^2 \tilde{\phi}_{\kx\ky}^2\tilde{\phi}_{\ky}  ,
    \nonumber
    \\
    \tilde{\phi}_{\kx\ky}\gg & \frac{\tilde{\mathcal{N}}}{\tilde{\omega}_{\kx \ky}^\mathrm{lin}} \tilde{\phi}_{\ky}\tilde{\phi}_{\kx}\sim \frac{\tilde{\mathcal{N}}^2}{\tilde{\omega}_{\ky}^\mathrm{lin}\tilde{\omega}_{\kx \ky}^\mathrm{lin}} \tilde{\phi}_{\ky}^2\tilde{\phi}_{\kx\ky} ,
    \nonumber
\end{align}
\end{subequations}
from which it follows that,
\begin{equation}
    \tilde{\phi}_{\kx},\tilde{\phi}_{\kx\ky},\tilde{\phi}_{\ky} \ll \Phi_\mathrm{crit}\stackrel{\cdot}{=} \frac{\tilde{\omega}_{\ky}^\mathrm{lin}}{\tilde{\mathcal{N}}} ,
\end{equation}
where $\Phi_\mathrm{crit}$, resulting from the balance of the linear time scale of the primary and the nonlinear interaction, sets a critical perturbation magnitude below which the weak regime is valid. As in the linear regime, this inequality cannot be satisfied \textit{ad infinitum}, and at some time $t_1$ the growing $\tilde{\phi}$ will reach the constant $\Phi_\mathrm{crit}$ threshold. The weakly regime will be readily visible so long as the orderings can be consistently satisfied over a time window $\Delta t\sim1/\omega_{\ky}^\mathrm{lin}$. 

\subsection{Strongly Nonlinear Regime}\label{sec:intro_strong}
We now consider the immediate extension to the above, in which both the zonal and sideband modes evolve nonlinearly, while the primary mode is the only one to remain linear. The minimal ingredients for defining the strongly nonlinear regime are
\begin{align}
	\begin{cases}
        \displaystyle
		\mathcal{N}_{\ky} & = 0, \\
        \displaystyle
        \mathcal{N}_{\kx \ky} & = ic \kx \ky
		\left[ J_{0,\kx}\,\phi_{\kx}\,\dgg_{\ky} - J_{0,\ky}\,\phi_{\ky}\,\dgg_{\kx} \right], \\
		\displaystyle
		\mathcal{N}_{\kx} & = ic \kx \ky \left[J_{0,\ky}\,\phi_{-\ky}\,\dgg_{\kx \ky}- J_{0,\kx \ky}\,\phi_{\kx \ky}\,\dgg_{-\ky} \right] .
	\end{cases}
    \label{eq:strongly_nl_terms}
\end{align}
To assess this regime we must once again consider the orderings. The characteristic time scales of the zonal and sidebands are determined by consistently satisfying $\omega_{\kx}^\mathrm{NL}\approx\omega_{\kx\ky}^\mathrm{NL}-(\omega_{\ky}^\mathrm{lin})^*$ and $\omega_{\kx\ky}^\mathrm{NL}\approx\omega_{\kx}^\mathrm{NL}+\omega_{\ky}^\mathrm{lin}$. Taking the nonlinear complex frequencies and growth rates to be comparable, $\omega\sim\gamma$, the essential requirement is that the zonal and sideband dynamics occur on a time scale much faster than that of the primary mode
\begin{equation}
    \gamma_{\ky}^\mathrm{lin} \ll \omega_{\kx} \sim \omega_{\kx \ky}. 
    \label{eq:strongly_growth_order}
\end{equation}
Following from this, it can be shown that the zonal and sideband amplitudes should be comparable, $\tilde{\phi}_{\kx}\sim\tilde{\phi}_{\kx\ky}$. 

For the primary to remain linear, we insist that over a time window of interest, $\Delta t\sim1/\gamma_{\kx}$, the nonlinear correction to the primary must remain small. Estimating this contribution by integrating the nonlinear term over $\Delta t$, we obtain the scaling requirement
\begin{equation}
    \tilde{\phi}_{\ky}\gg \frac{\tilde{\mathcal{N}}}{\tilde{\gamma}_{\kx}^\mathrm{NL}} \tilde{\phi}_{\kx\ky}\tilde{\phi}_{\kx}.
    \label{eq:strong_primary_amp_ordering}
\end{equation}  
On the other hand, for the zonal and sideband equations to exhibit dominantly nonlinear behavior, the primary must be sufficiently strong. This yields the constraints
\begin{equation}
    \tilde{\phi}_{\ky}\gtrsim \frac{\tilde{\omega}_{\kx}^\mathrm{lin}}{\tilde{\mathcal{N}}}\frac{\tilde{\phi}_{\kx}}{\tilde{\phi}_{\kx\ky}}, \frac{\tilde{\omega}_{\kx\ky}^\mathrm{lin}}{\tilde{\mathcal{N}}}\frac{\tilde{\phi}_{\kx\ky}}{\tilde{\phi}_{\kx}}.
    \label{eq:inequality_strong}
\end{equation}
This scenario set by Eq.~\eqref{eq:strongly_growth_order}, corresponds to the standard secondary–instability set-up in which the primary (streamer) magnitude is frozen in time. In this regime fall both the treatments of RDK \citep{Rogers2000} and of  \cite{Richard2024}. If, like RDK, one neglects geometric effects from the linear contributions to the zonal and sideband equations, one finds
\begin{equation}
\tilde{\omega}_{\kx}^\mathrm{NL}\sim \tilde{\mathcal{N}} |\tilde{\phi}_{\ky}|.
\end{equation}
This corresponds to the celebrated super-exponential RDK secondary growth of the ZF. To retain consistency with the condition in Eq.~\eqref{eq:strong_primary_amp_ordering}, the primary must be sufficiently strong, $\tilde{\phi}_{\ky}^2\gg \tilde{\phi}_{\kx}\tilde{\phi}_{\kx\ky}$, and must be so over a sufficiently long time window. We present in Appendix~\ref{sec:RDK_appendix} a more detailed comparison to the original gyrofluid work by RDK.

Including the geometry for the ZF and sideband brings us closer to the work of \cite{Richard2024}. In Appendix~\ref{sec:toroidal_secondary_appendix} we prove the equivalence between the original work, and our truncated modal approach, under the same subsidiary ordering $\ky \gg \kx$. A crucial point in establishing this equivalence is the need to include an additional mirror mode to the minimal triadic model; namely, the mode at finite $\kx$ and negative but finite $\ky$. We must correspondingly add an extension to Eqs.~\eqref{eq:strongly_nl_terms} (see Figure~\ref{fig:strongly_nl_minimal_mode} and details in Appendix~\ref{sec:toroidal_secondary_appendix}). 
\subsection{Fully Nonlinear Regime}\label{sec:intro_full}
Finally, in the fully nonlinear regime, all modes evolve according to their nonlinear GK equations. The corresponding nonlinear terms are, for completeness, given by
\begin{align}
	\begin{cases}
        \displaystyle
        \mathcal{N}_{\ky} & = ic \kx \ky \left[ J_{0,\kx \ky}\,\phi_{\kx \ky}\,\dgg_{-\kx} - J_{0,\kx}\,\phi_{-\kx}\,\dgg_{\kx \ky} \right], \\
        \displaystyle
		\mathcal{N}_{\kx \ky} & = ic \kx \ky \left[ J_{0,\kx}\,\phi_{\kx}\,\dgg_{\ky} - J_{0,\ky}\,\phi_{\ky}\,\dgg_{\kx} \right], \\
		\displaystyle
		\mathcal{N}_{\kx} & = ic \kx \ky \left[ J_{0,\ky}\,\phi_{-\ky}\,\dgg_{\kx \ky} - J_{0,\kx \ky}\,\phi_{\kx \ky}\,\dgg_{-\ky}
		\right].
	\end{cases}
\end{align}

\section{Weakly Nonlinear Regime}\label{sec:weaklyNL}
For the remainder of this paper we shall focus on ZF evolution in the weakly nonlinear regime, described in Section~\ref{sec:intro_weak}. The most straightforward way to extract the dominant behaviour of the ZF in this regime is to treat the problem as a forced system. For this, we first invoke the triadic model, which we repeat here for clarity 
\begin{align}
    \mathcal{L}_{\vec{k}} \, \dgg_{\vec{k}} = \mathcal{R}_{\vec{k}}  \frac{q \maxwellian}{T_{0}} J_{0, \vec{k} } \phi_{\vec{k}} \quad &\mbox{for} \quad \vec{k} =
	\begin{cases}
        \displaystyle
        (0,\bky)
        \\
        \displaystyle
		(\bkx, \bky),
	\end{cases}
    \label{eq:nonzonal}
    \\
    \mathcal{L}_{\vec{k}} \, \dgg_{\vec{k}} = \mathcal{R}_{\vec{k}} \frac{q \maxwellian}{T_{0}} J_{0, {\vec{k}} } \phi_{\vec{k}} - \mathcal{N}_{\vec{k}} \quad &\mbox{for} \quad \vec{k} = (0, \bkx) .
    \label{eq:zonal}
\end{align}
We further assume that Eq.~\eqref{eq:nonzonal} admits a single unstable normal mode solution, $\phi_{\ky} \sim \exp(-i\omega_{\ky} t) $, and $\phi_{\kx \ky} \sim \exp(-i \omega_{\kx \ky} t) $. From wavenumber matching, the temporal behaviour of $\mathcal{N}_{\bkx}$ is
\begin{equation}
    \omega_{\bkx} = \omega_{\bkx,\bky} - \omega_{\bky}^* ,
    \label{eq:2gamma_eqn}
\end{equation}
which then drives the ZF at that same frequency. Assuming that the primary and sideband both behave similarly with growth rate $\gamma_{\bky}$ (which is the case in the limit of $\bkx \rightarrow 0$, or $\omega_{\bkx, \bky} \sim \omega_{\bky}\gg \omega_{d, \kx}$) this gives
\begin{equation}
    \omega_{\bkx} =\omega_{\bkx \bky}-\omega_{\bky}^* \approx 2i\gamma_{\bky}.
    \label{eq:2gamma_smallkx}
\end{equation}
Thus, for strong nonlinear drive, and $\kx\ll \ky,$ the zonal flow is bound to be purely growing, and to grow with a rate twice that of the primary mode \citep{todo2010nonlinear}. This, in fact, is a special case of the convolution condition in Eq.~\eqref{eq:2gamma_eqn}, which has been verified numerically. Note that this is a purely driven response, with no marginal points, and largely independent of the initial ZF condition. Consequently, provided the primary mode is unstable, this behaviour will inevitably be observed if the system is evolved for a sufficiently long time. Generally, as one goes to higher $\kx \rho_i$ the sideband will become linearly stabilised, and thus this mechanism will become subdominant. One should therefore frame much of this analysis to be primarily relevant for $\kx \rho_i < 1$.

\subsection{Numerical Justification}
We substantiate the triadic model in this regime through a set of numerical experiments using the electrostatic version of the $\delta \! f$-GK code, \texttt{stella} \citep{Barnes19}, with a modified adiabatic electron response, and Cyclone Base Case (CBC) geometry. These, and all other numerical results, are presented in terms of the code-variables, $k_x |\nabla x| = \kx |\nabla \psi |$ and $k_y |\nabla y| = k_{\alpha} |\nabla \alpha |$ (see Appendix~\ref{sec:numerical_studies} for details on all numerical set-ups in this paper). 

\begin{figure}
    \centering
    \includegraphics[width=\textwidth]{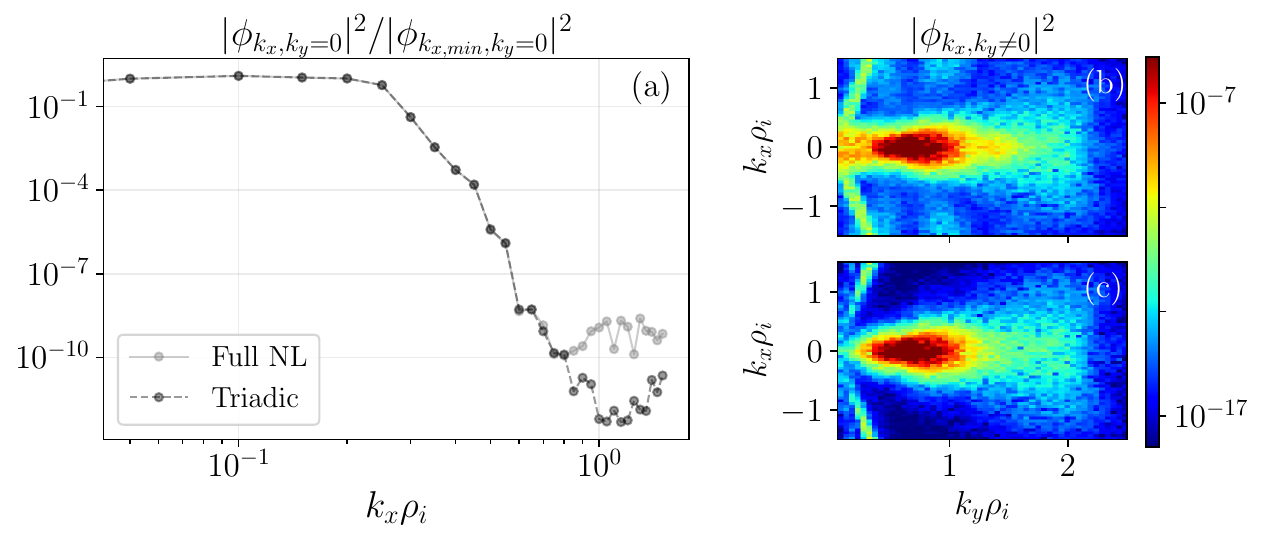}
    \captionof{figure}{\textbf{Validation of the Weakly Nonlinear Regime.} Comparison of $|\phi_{k_{x}, k_{y}}|^2$ spectra from a fully nonlinear, and artificially imposed weakly nonlinear simulation. The latter corresponds to the system described by Eqs.~\eqref{eq:nonzonal}-\eqref{eq:zonal}. Figure (a) shows a comparison of the zonal components as a function of $k_x \rho_i$, whilst (b) and (c) show the spectra of the non-zonal components.}
    \label{fig:phi2_spectra}
\end{figure}

As initial step we verify the fundamental ingredient of the weakly nonlinear regime, namely that only the ZF evolves nonlinearly. To do this, we conduct a simulation in which we artificially turn off the non-zonal nonlinearities, and compare this to a fully nonlinear GK simulation. The results are presented in Figure~\ref{fig:phi2_spectra}. They show that the zonal spectrum is well captured by the weakly nonlinear assumption up to $k_x \rho_i \sim 1$. This is similarly true for the non-zonal spectra, and although differences are observable many of the key features remain. 

We must further check the validity of the triadic assumption. For this we first need a numerical way to identify a triad for a given $k_x$. We determine the triad by considering each $k_x$ in turn and identifying its corresponding sideband. Specifically, at fixed $k_x$, the linear growth rate is computed as a function of $k_y$, and the value of $k_y$ at which the growth rate is maximised is then selected for that triad.

\begin{figure}
    \centering
    \includegraphics[width=\textwidth]{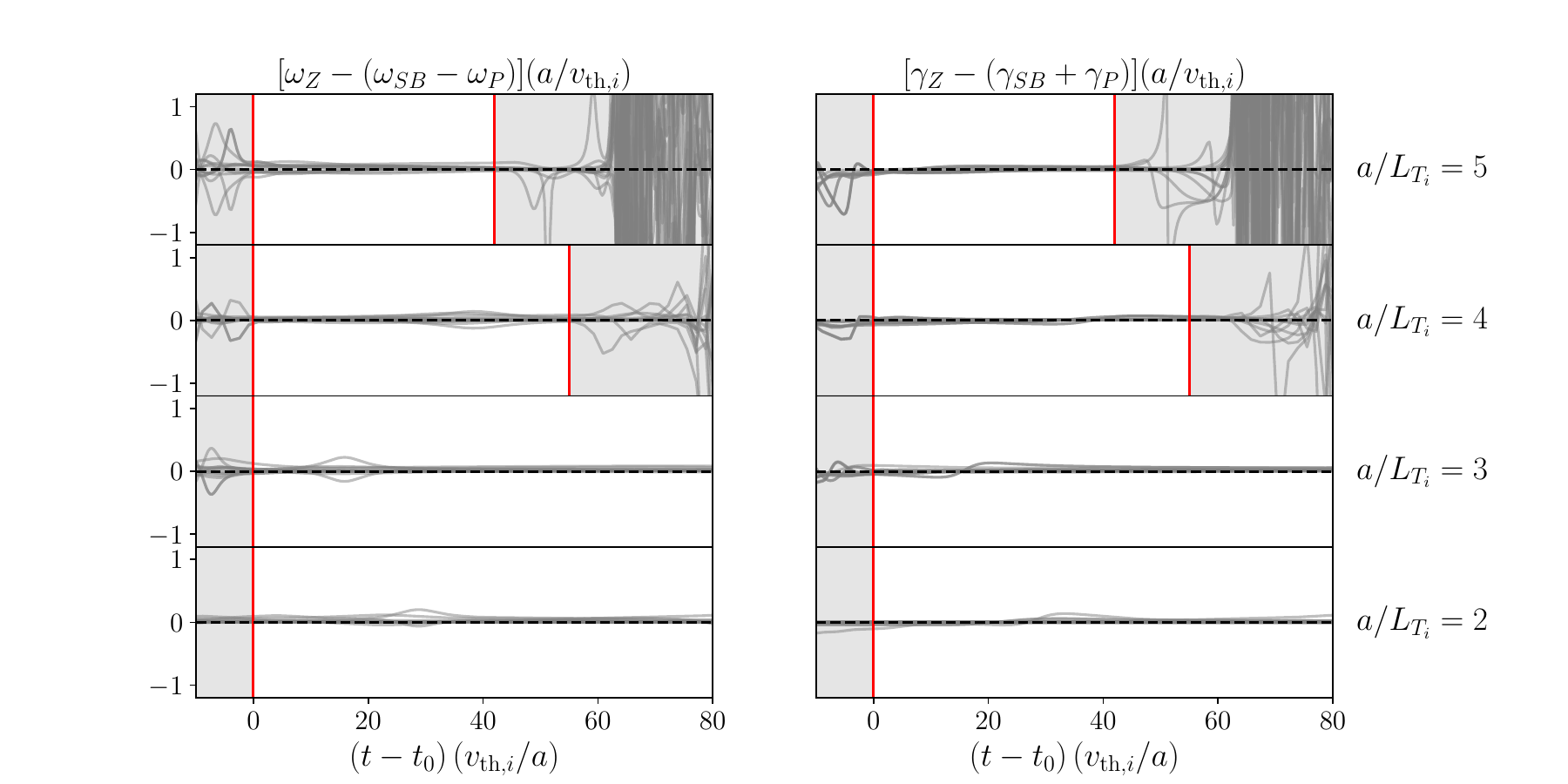}
    \caption{\textbf{Validation of the Triadic Assumption.} Comparison of the frequency (left) and growth rate (right) of the zonal mode obtained from nonlinear simulations with the values predicted by the convolution of linear modes, Eq.~\eqref{eq:2gamma_eqn}. In each case, the onset of the weakly nonlinear regime, here denoted as $t_0$, and the breakdown into the strongly/fully noninear regime is indicated by the red lines. From top to bottom, the normalised temperature gradients of the simulations are $a/ L_{T_i} = 5, 4, 3$, and $2$.}
    \label{fig:convolution}
\end{figure}

Once individual triads have been identified\footnote{For simplicity, each triad is found using the linear $k_y$ spectra for a given $k_x$ for the $a/L_{T_i} = 5$ case.} in the range $k_x \rho_i \in [-0.3, 0.3]$, we test the convolution in Eq.~\eqref{eq:2gamma_eqn}, analysing full nonlinear GK simulations with different temperature gradients. We choose this window in $k_x \rho_i$ based on the linear instability of the sideband, which is weak outside of the range considered. The results are presented in Figure~\ref{fig:convolution}. The comparison of the frequency and growth rates of the ZF matches the values predicted by the convolution. It is evident from this that although in reality multiple triads for each $k_x$ exist, assuming just a single dominant triad suffices. We note that the agreement only holds over a finite time interval in which the weakly nonlinear regime is valid, after which the system falls into the strongly/fully nonlinear regimes.

\begin{figure}
    \centering
    \includegraphics[width=0.6\textwidth]{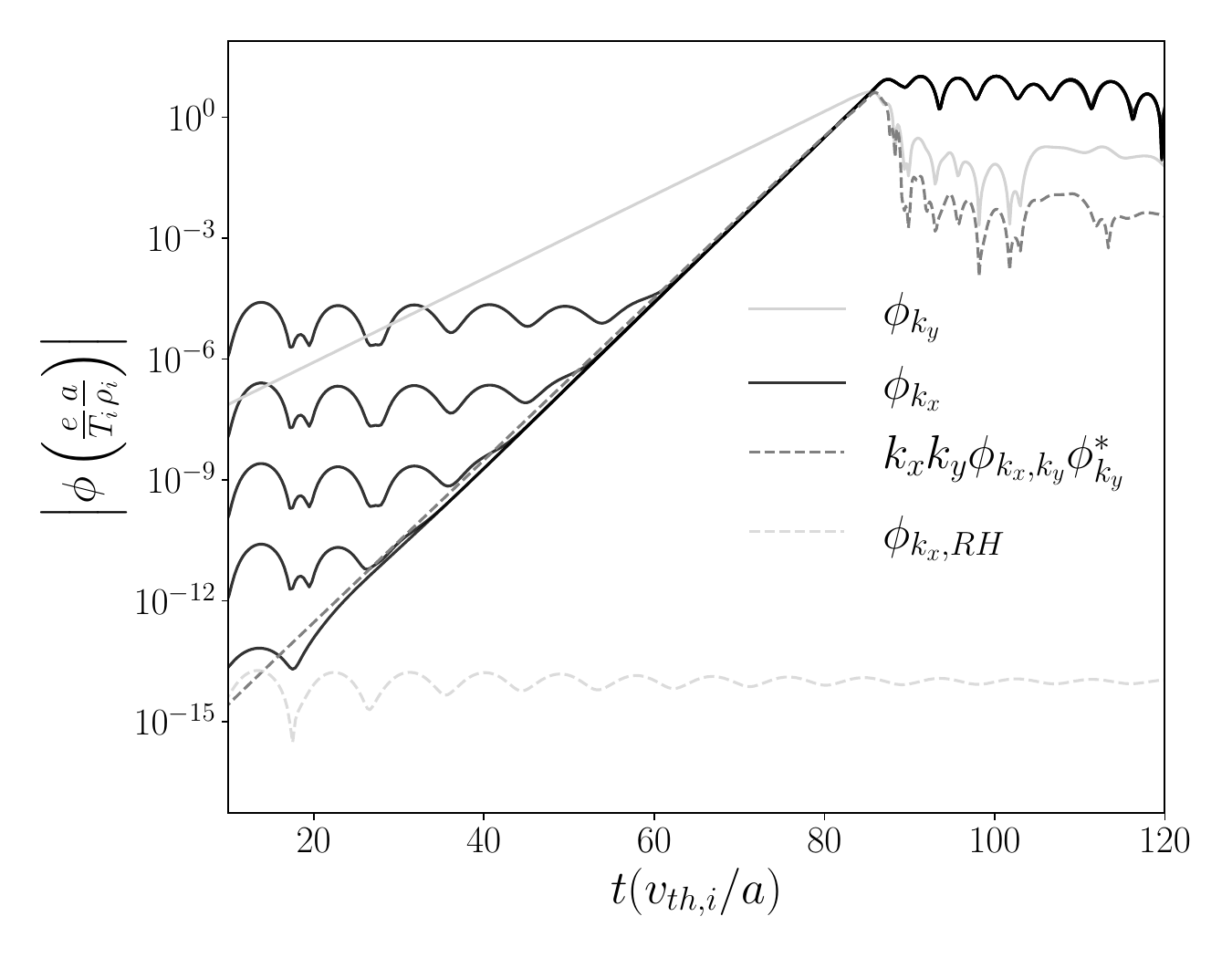}
    \captionof{figure}{\textbf{Simulations of Rescaled Initial ZF.} Evolution of the zonal potential for simulations initialised with different zonal amplitudes. In each simulation, the non-zonal modes are initialised the same way, and the phase space structure of the zonal is maintained. Also shown is a proxy for the zonal response, proportional to $k_x k_y\,\phi_{k_xk_y} \phi_{k_y}^*$, given by the dashed line. The purely linear zonal response is plotted in light grey for reference, $\phi_{Z,RH}$.
    }
    \label{fig:different_initial_conditions}
\end{figure}

The final consideration is to justify the concept of a driven system. For this, we wish to show that, when dominant, the behaviour in the weakly nonlinear regime is independent of the initial amplitude of the ZF. To demonstrate this we conduct a numerical experiment in which the zonal mode is initialised with the same phase-space structure, but with its amplitude rescaled at $t=0$. All other modes in the system are initialised the same across the different simulations. The time traces of a selected $k_x$-mode are shown in Figure~\ref{fig:different_initial_conditions}, alongside a proxy for its nonlinearity, $k_x k_y \phi_{k_x k_y}\phi_{k_y}^*$\footnote{The choice to focus on a single $k_x$ mode is made purely for illustrative purposes; the same qualitative behaviour is obtained for other values of $k_x$.}. In all rescaled cases, the behaviour in the weakly nonlinearly regime, and thereafter, is identical. Prior to this, the nonlinear driving is still present yet not observable because it is subdominant.

\subsection{Analysis of Weakly Nonlinear Regime}\label{sec:Laplace}

Given the simple set-up in the weakly nonlinear regime, it is possible to directly analyse the structure of the ZF as it is driven. We derive this structure assuming the dominant Fourier analysis considered at the beginning of Section~\ref{sec:weaklyNL}. For this, we start with the quasineutrality condition for the zonal component of the distribution function
\begin{equation}
    \frac{1}{n_{0}}\int \rmd^{3}\mathbf{v}J_{0, \bkx} \dgg_{\bkx}= \left(\frac{1}{\tau}+1\right)\frac{e\phi_{\bkx}}{T_{0e}} - \frac{e}{T_{0e}}\frac{1}{\tau} \langle \phi_{\bkx} \rangle_{\psi},
\end{equation}
where we have taken the electrons to have a modified adiabatic response, $J_{0,e,\veck} \sim 1$, and $\tau = T_{0,i}/T_{0,e}$.
We then use the nonlinear Eq.~\eqref{eq:zonal} for $h_{\bkx}$, and the linear Eqs.~\eqref{eq:nonzonal} for $h_{\bky}$ and $h_{\bkx\bky}$ (dropping the streaming terms), to yield the following amplitude for the ZF, $\phi_{\bkx}^{\omega_{\bkx}}$, 
\begin{equation}
    \epsilon_{\bkx} \phi_{\bkx}^{\omega_{\bkx}} = -  i \frac{\omega_{\mathbf{E}\times\mathbf{B}} }{\Omega_{\mathrm{NL}}} \phi_{-\bky}^{\omega_{\bky}}, 
    \label{eq:zonal_qn_deltafn}
\end{equation}
with
\begin{subequations}
\begin{align}
    \epsilon_{\bkx} =& \int \rmd^{3}\mathbf{v}\frac{F_{0i}}{n_{0}}J_{0, \bkx}^{2} \frac{\omega_{\bkx}}{\omega_{\bkx}-\hatomegadx}-1 ,
    \label{eq:epsilon_fourier}
    \\
    \Omega_\mathrm{NL}^{-1}=&-\int \rmd^{3}\mathbf{v}  \frac{F_{0i}}{n_{0}}
    \frac{J_{0, \bkx} J_{0, \bky}J_{0, \bkx\bky}}{\omega_{\bkx}-\hatomegadx} \mathcal{I}
    \label{eq:omega_nl}
    \\
    \mathcal{I} = &
        \frac{\omega_{\bkx\bky}-\hat{\omega}_{*, \bky}}{\omega_{\bkx\bky}-\hatomegadxy} 
        -
        \frac{\omega_{-\bky}-\hat{\omega}_{*,-\bky}} {\omega_{-\bky}-\hatomegadmy}
    ,
    \label{eq:I_fourier}
\end{align}
\end{subequations}
where $\omega_{\mathbf{E}\times\mathbf{B}}= c\bkx\bky\phi_{\bky}^{\omega_{\bkx\bky}}
$, $\hat{\omega}_{*, \ky} = \omega_{*, \ky} \left[ 1 + \eta (\hat{v}^2 - 3/2) \right]$, $\eta = \rmd(\ln T) / \rmd (\ln n)$, $\hat{\omega}_{d, \lambda} = \omega_{d, \lambda} \left(\hat{v}_{\parallel}^{2}+ \hat{v}_{\perp}^{2}/2 \right)$, and $\hat{v} = v / \vths $ indicates the normalised velocity coordinate. Here $\epsilon_{\bkx}$ represents the linear zonal response, whilst $\Omega_\mathrm{NL}^{-1}$ expresses the nonlinear driving. It should be stressed that Eq.~\eqref{eq:zonal_qn_deltafn} should not be interpreted as a dispersion relation, but instead as an equation for the amplitude of $\phi_{\bkx}^{\omega_{\bkx}}$ in terms of linear primary amplitudes, with $\omega_{\bkx}$ already determined by the complex frequency matching from Eq.~\eqref{eq:2gamma_eqn}.

A more careful approach employing the Laplace transform yields the same dominant behaviour provided the non-zonal modes are strongly unstable (which is generally the case for small $\kx \rho_i$). The Laplace approach helps setting up the necessary integrals involved in quasineutrality properly, and is more careful to not disregard any potentially important time response of the ZF. This analysis is performed in Appendix~\ref{sec:laplace_derivaion}, demonstrating the lack of marginal stability point and highlighting the fundamentally driven nature of the system (see Figure~\ref{fig:different_initial_conditions}).

The nonlinear integral in Eq.~\eqref{eq:omega_nl} can be written in terms of more manageable functions 
\begin{equation}
    \Omega_{NL}^{-1}=\frac{1}{\omegadx \omega_{\bky}^{*} - \omega \omegady} \left[
    \omegadxy I_{\bkx \bky}^{\omega_{\bkx \bky}} 
    - \omegadx I_{\bkx}^{\omega_{\bkx}}
    - \omegady I_{-\bky}^{-\omega_{\bky}^{*}}
    \right],
    \label{eq:NLresonance}
\end{equation}
where we define,
\begin{equation}
    I_{\beta}^{\omega} \left( \omega_{*,\ky} \right) = \int \rmd^{3}\mathbf{v} \frac{F_{0i}}{n_{0}} \frac{\omega - \hat{\omega}_{*, \ky}^T}{\omega - \hat{\omega}_{d,\beta}} \besseltrio.
    \label{eq:integral_definitions}
\end{equation}
These integrals are derived in Appendix \ref{sec:resonant_nl_integral}, and a closed form in terms of plasma dispersion functions presented when dropping FLR effects \citep{Biglari89,zocco2018,ivanov2023analytical}. In that limit $\epsilon_{\bkx} = I_{\bkx}^{\omega_{\bkx}} - 1 $ also holds.


To further understand this amplitude response and its implications, let us now consider the limit for which $\kx \rightarrow 0$. 
Retaining the FLR effects and expanding both sides in $\omega_{d, \bkx}/\omega_{\bkx} \sim \bkx |\nabla \psi| \rho_{i} \ll 1$ and $\omega_{d, \bky}/\omega_{\bky} \sim \bky |\nabla \alpha| \rho_{i} \ll 1$, with $\omega_{\bkx}\sim \omega_{\bky}$, we find 
\begin{equation}
    \epsilon_{\kx} \approx \frac{\omega_{d,\bkx}}{\omega_{\bkx}}- b_{\bkx},
    \label{eq:zonal_linear_smallkx}
\end{equation}
and
\begin{multline}
    \Omega_{\text{NL}}^{-1}\approx \frac{1}{\omega_{\bky}^*(\omega_{\bky}^*+ \omega_{\bkx})\omega_{\kx}} 
    \left[\frac{1}{2} \left( b_{\bkx} + b_{\bky} + b_{\bkx, \bky } \right) 
    \omega_{*,\ky}^T \omega_{\bkx}
    \right.
    \\
    \left.
    + \omega_{d,\bky}\omega_{\bkx}  - \omega_{d,\bkx} \omega_{\bky}^* 
    - \frac{\omega_*^T}{\omega_{\bky}^{*}} 
    \frac{ \omega_{d,\bkx} \omega_{\bky}^* \omega_{\bkx} 
    +
    \omega_{d,\bky} (2\omega_{\bky}^*\omega_{\bkx} + \omega_{\bkx}^2) }{(\omega_{\bky}^* + \omega_{\bkx})}
    \right].
    \label{eq:nl_smallkx}
\end{multline}
Note that all terms in Eqs.~\eqref{eq:zonal_linear_smallkx} and \eqref{eq:nl_smallkx} are proportional to either $b_{\vec{k}}$ or $\omega_{d,\vec{k}}$. The dependence of $\epsilon_{\bkx}$ in particular results from the compression response of the zonal arising from either geodesic curvature or FLR effects.

In the long wavelength limit with finite geodesic curvature, this shows that one always gets a finite zonal response. Further assuming $\Re(\omega_{\vec{k}} ) \ll \Im(\omega_{\vec{k}})$,
\begin{equation}
    \phi_{\bkx}^{\omega_{\bkx}} 
    \approx 2i
    \frac{c}{\gamma/\bky}\frac{\omega_{d, \bky}/\bky}{\omega_{d, \bkx}/\bkx} \bky \phi_{\bkx, \bky}^{\omega_{\bkx\bky}} \left(\phi_{\bky}^{\omega_{\bky}}\right)^*  
\end{equation}
to leading order, $\mathcal{O}(1)$. Assuming that $\gamma \propto \bky$, the zonal amplitude is larger when the primary has a shorter length scale, as is shown by the linear dependence on $\bky$. The geometry of the field also plays an important role in controlling the strength of the zonal response. Both the normal and geodesic curvatures are involved, balancing each other to give a finite response at long wavelengths. Configurations with a reduced geodesic curvature will therefore see a very strong growth of large scale zonal-flows. This is an enhancement that aligns with the geometric considerations following the ZF residual in stellarators \citep{plunk2024residual,Rodriguez_Plunk_2025}.

This behaviour invites us to consider the limit of vanishing geodesic curvature, as is the case in a $Z$-pinch ($\omega_{d,\kx} \rightarrow 0$ and $\bkx \rho_i \ll 1$). In this limit we find

\begin{equation}
    \phi_{\bkx}^{\omega_{\bkx}} 
    \approx 
    - i \frac{c}{\bkx} \phi_{\bkx, \bky}^{\omega_{\bkx \bky}} \left(\phi_{\bky}^{\omega_{\bky}}\right)^{*}\frac{1}{(\gamma/\bky)^2}\left[ (b_{\bkx} + b_{\bky} + b_{\bkx, \bky}) \omega_{*,\bky}^T/\bky + 2\omega_{d, \bky}/\bky \right] \propto \frac{1}{\bkx}.
\end{equation}
That is, the zonal amplitude is large (in fact, diverges) for small $\bkx$. This suggests that in geometries with small but finite geodesic curvature there will be a rapid growth of large-scale zonal modes through the $2\gamma$-mechanism that leads the system to quickly leave the weakly nonlinear regime and transition into a strongly nonlinear or fully saturated state.



\section{Appearance of a Large Residual}\label{sec:res_2_gamma}
The discussion about the zonal response in the weakly nonlinear regime thus far has focused on the electrostatic potential amplitude. Little was said about its velocity-space structure, which is however important to fully understand the nature of the ZF. In particular, the level of the ZF residual is known to depend on the precise form of its distribution function \citep{monreal2016residual}. In this section we study this.

Let us start our study of the phase-space structure of the zonal mode by considering a simplified zonal distribution function. As will be shown later, this structure naturally arises in the limit of a strongly driven toroidal ion-temperature gradient (ITG), in which the drift contributions to the ZF can be neglected, $\omega_{*}\gg \omega_{d,\vec{k}}, \omega_{k_x}$ (see e.g.\ \cite{kadomtsev2012reviews, Biglari89}). Assuming a dominant temperature gradient ($\eta\gg1$), such that $h_{\bky} \sim q/T J_{0,\bky} \hat{\omega}_{*,\bky}/\omega_{\bky} F_0 \phi_{\bky} $ from Eq.~\eqref{eq:nonzonal}, Eq.~\eqref{eq:zonal} yields,
\begin{subequations}\label{eq:gen_weak_dF_full}
    \begin{gather}
        \langle \delta f_{\bkx}\rangle_\mathbf{R} = \hat{\mathcal{N}} J_{0, \bky}J_{0, \bkx \bky} W_N F_0,
        \label{eq:gen_weak_dF}
        \intertext{with}
        \hat{\mathcal{N}} = -i \bky \bkx c\frac{q}{T}\phi_{\bky}^*\phi_{\bkx}\frac{\omega_{*,\bky}^{T}}{\omega_{\bky}^*\omega_{\bkx\bky}}, \quad W_N=\hat{v}^2-\frac{3}{2}.
        \label{eq:ITG_structure_simple}
    \end{gather}
\end{subequations}
The velocity-space dependence is inherited from that of the diamagnetic frequency, explicitly $\hat{\omega}_{*,\bky}= \omega_{*,\bky}^{T}(\hat{v}^2-3/2)$. This resulting velocity structure leads to the perturbed distribution only weakly contributing to the perturbed electrostatic potential, but strongly to the temperature perturbation. We can see this explicitly, 
\begin{subequations}
\begin{align}
    \frac{\delta n}{n} \sim &\frac{1}{n}\int\rmd^3\mathbf{v}J_{0 \kx} \langle\df_{\kx} \rangle_\mathbf{R}= -\frac{\hat{\mathcal{N}}}{2}(b_{\kx}+b_{\kx\ky}+ b_{\ky}) 
    \label{eq:dn} \\
    \frac{\delta T}{T} \sim &\frac{1}{nT}\int\rmd^3\mathbf{v}J_{0 \kx}v^2 \langle\df_{\kx} \rangle_\mathbf{R}= \frac{3}{2}\hat{\mathcal{N}}, 
    \label{eq:dT}
\end{align}
\end{subequations}
where we have ignored any $z$-dependence, expanded in the smallness of $k_\perp \rho_i$, and considered the perturbed density and temperatures at fixed guiding centers. For this, we have made direct use of Weber integrals as given in \cite{gradshteyn2014table}(6.614.1 and 6.615) and $b_{k_{\lambda}}=(k_{\lambda} |\nabla \lambda| \rho_i)^2/2$. This shows that most of $\df$ averages out in velocity space. 

\begin{figure}
    \centering
    \includegraphics[width=0.8\textwidth]{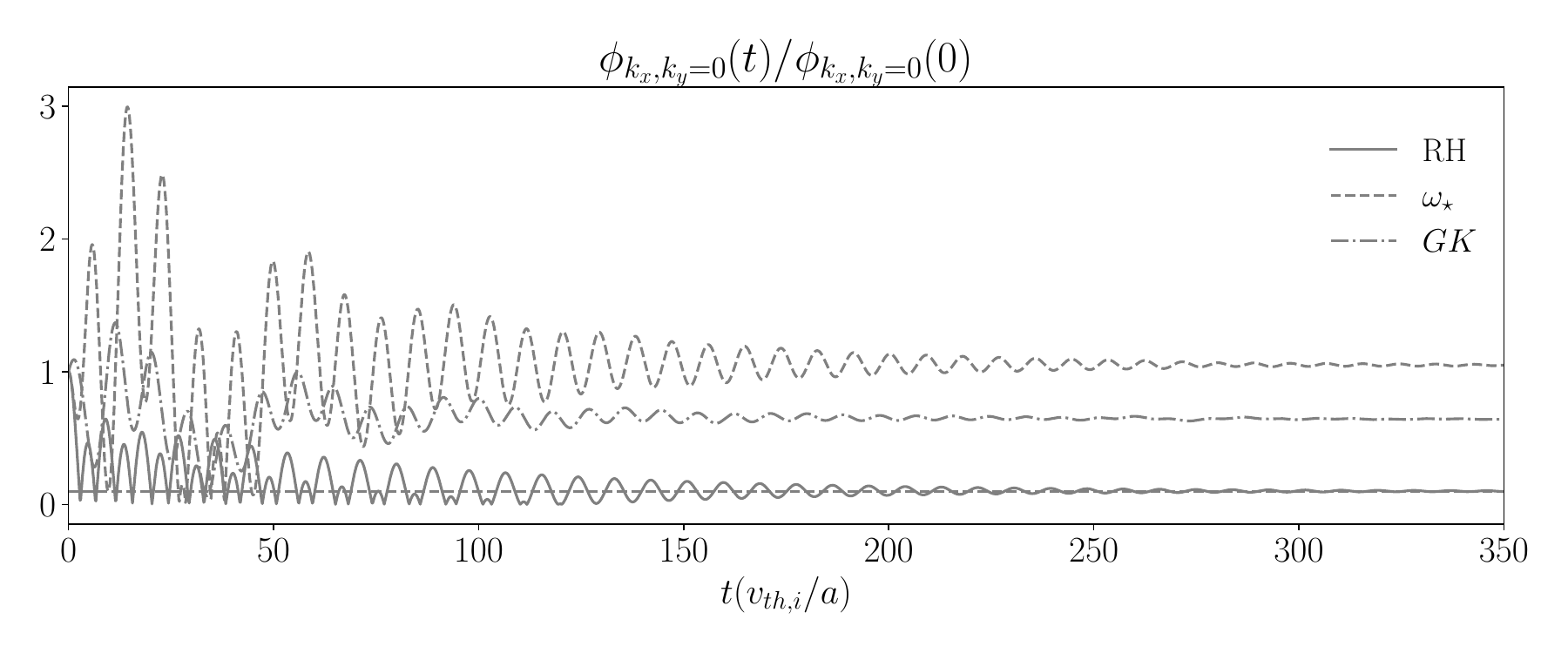}
    \caption{\textbf{Rosenbluth-Hinton Test for Different Initial Zonal Distributions.} Time traces of zonal modes for three ZF initial conditions with identical potential amplitude, but different velocity-space structure: the classical RH density perturbation, Eq.~\eqref{eq:RH_df_0}, the $\omega_*$ temperature perturbation, Eq.~\eqref{eq:gen_weak_dF}, and the initial condition taken from the gyrokinetic simulation in the $2\gamma$ regime. All potentials are normalised to unity at $t=0$ and evolved using the full linear gyrokinetic equation.}
    \label{fig:residual_compare_time_trace}
\end{figure}

This form of the ZF is very different to what one may customarily consider; \textit{i.e.}, from a ZF dominated by a density perturbation. So will basic questions about it, such as the linear Rosenbluth \& Hinton (RH) test, which probes its resilience by evolving it linearly for an infinite time. 
In Figure~\ref{fig:residual_compare_time_trace} we compare the time traces of such test for three different cases using \texttt{stella}. The cases show substantial variation: the classical RH initialisation yields a significantly lower residual compared to either the simple initial condition in Eq.~\eqref{eq:gen_weak_dF}, or that taken directly from the nonlinear gyrokinetic simulation of the weakly nonlinear regime. 

To make sense of these differences, the residual value can be found quite generally for any given initial $\df$ \citep{monreal2016residual}. Assuming $\phi_\infty$ (the ZF amplitude in the limit of $t\rightarrow \infty$ when evolved linearly) to be independent of the fieldline-following coordinate, and defining $Q$ as the finite orbit width solution to the equation $v_\parallel\nabla_\parallel Q=\omega_{d, \kx}$, 
\begin{subequations}\label{eq:normalised_residual}
    \begin{align}
        \langle\phi_\infty\rangle_\psi=
        \frac{\frac{1}{n}\left\langle \int\rmd^3\mathbf{v}J_{0,\kx}e^{-iQ}\overline{e^{iQ}\langle \df_{\kx,t=0}\rangle_\mathbf{R}} \right\rangle_\psi}{1-\frac{1}{n}\left\langle \int\rmd^3\mathbf{v}J_{0, \kx}e^{-iQ}\overline{J_{0,\kx} e^{iQ}} \maxwellian\right\rangle_\psi}, \label{eq:phi_infty} \\
        \langle\phi_0\rangle_\psi=\frac{1}{1-\Gamma_0}\frac{1}{n}\left\langle\int\rmd^3\mathbf{v}J_{0,\kx} \langle\df_{\kx,t=0}\rangle_\mathbf{R} \right\rangle_\psi,
    \end{align}
\end{subequations}
where the overline denotes bounce average and we considered omnigeneous fields. This exercise is generally performed with a pure initial density perturbation given, namely
\begin{equation}
    \langle\df_{\kx}\rangle_\mathbf{R}^\mathrm{RH}=b_{\kx}F_0\frac{q\phi_0}{T}, \label{eq:RH_df_0}
\end{equation}
where $\phi_0$ is the initial value of the electrostatic potential. When modelling the geometry as a large aspect ratio circular cross-section tokamak, the celebrated residual expression, is obtained 
\begin{equation}
    \left.\frac{\phi_\infty}{\phi_0}\right|_\mathrm{RH}=\frac{1}{1+\alpha_\mathrm{RH}},
\end{equation}
where $\alpha_\mathrm{RH}=1.6q^2/\sqrt{\Delta}$, $q$ is the safety factor and $\Delta$ the mirror ratio \citep{RosenbluthHinton1998,XiaoCatto2006}. 

\begin{figure}
    \centering
    \includegraphics[width=0.45\textwidth]{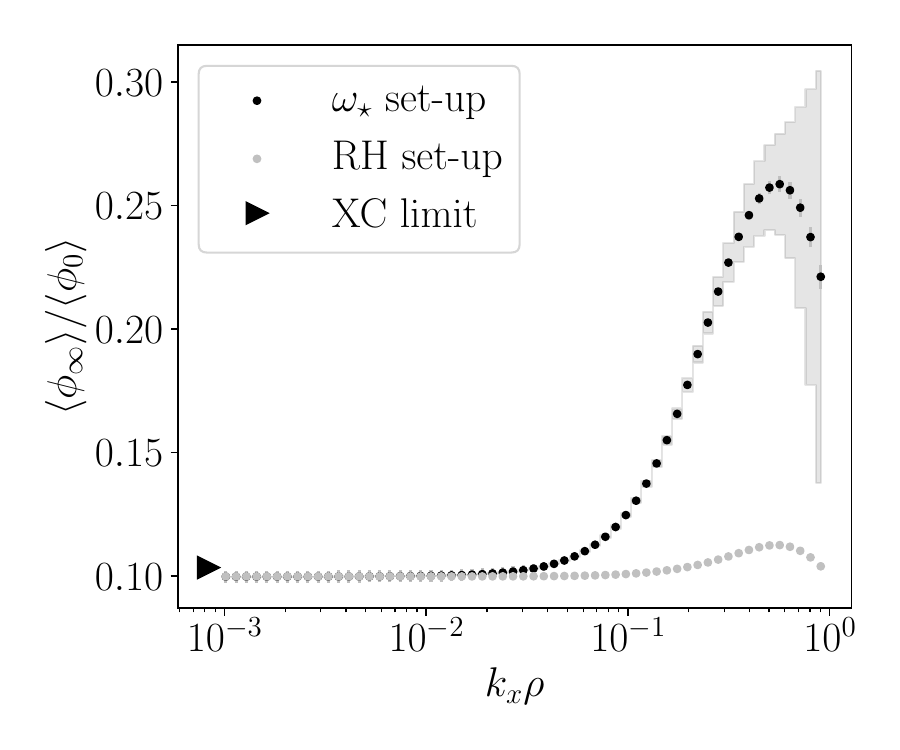}
    \includegraphics[width=0.45\textwidth]{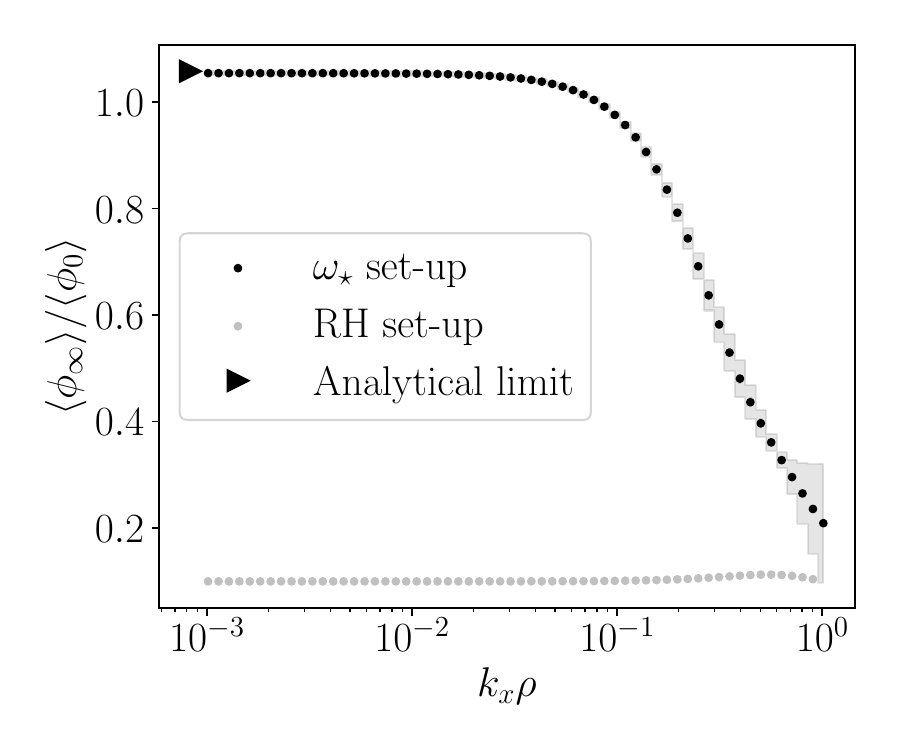}
    \caption{\textbf{Simplified Model Residual Level for Weakly Driven Regime.} Radial-wavenumber spectra of the zonal-mode residual for the simplified initialisation given in Eq.~\eqref{eq:gen_weak_dF_full} as computed with \texttt{stella}.
    The left panel shows the results including $\ky$ FLR effects, while the right one ignores them. The shaded region shows the standard deviation of the residual in $z$, and the vertical error bars denote the standard deviation in time. The triangular marker indicates the Xiao--Catto residual limit \citep{XiaoCatto2006}, and in gray the standard RH residual spectrum computed with \texttt{stella}. Details of the simulation are given in Appendix~\ref{sec:numerical_studies}.}
    \label{fig:residual_compare}
\end{figure}

The original work of Rosenbluth \& Hinton did also consider how an odd (in $v_\parallel$) part of an initial distribution would fare. However, they did not consider what we find to be, indeed, the most relevant component, namely, a significant $\delta T$ contribution. This can evolve into a potential yielding moment, as suggested by the derivation in Appendix~\ref{sec:ZF_residuals}, which considers the fluid moments of the distribution function.
More precisely, applying Eq.~\eqref{eq:normalised_residual} to the distribution function in Eq.~\eqref{eq:gen_weak_dF_full}, the resulting residual can be written explicitly in the limit of small $b_{\vec{k}}$ as
\begin{equation}
    \left.\frac{\phi_\infty}{\phi_0}\right|_\mathrm{\omega_*}=\frac{1+2\alpha_\mathrm{RH} b_{\kx}/(b_{\kx}+ b_{\ky} + b_{\kx \ky})}{1+\alpha_\mathrm{RH}}.
    \label{eq:residual_wstar}
\end{equation}
For further details, see Appendix~\ref{sec:ZF_residuals}. This shows indeed that the zonal potential is clearly more resilient compared to the typical RH one, $\phi_\infty/\phi_0|_{\omega_*}\geq \phi_\infty/\phi_0|_\mathrm{RH}$, as illustrated in the residual levels from gyrokinetic simulations of Figure~\ref{fig:residual_compare}. This difference tends to vanish for ZFs whose scale is much larger than that of the primary drive (see left plot of Figure~\ref{fig:residual_compare}). In that case, one may say that the linear evolution is incapable of re-purposing some of the distribution function for the potential. In the limit $\bky\leq\bkx$, the opposite is true, and there is an effective bound on the amplification effect on the residual (see right plot of Figure~\ref{fig:residual_compare}, where $b_{\bky} = 0$ is taken for Bessel functions). An estimate of the magnitude of this residual is obtained by taking the limit $b_{\bky} \rightarrow 0$ of Eq.~\eqref{eq:residual_wstar}, yielding $1 \leq \phi_\infty/\phi_0|_{\omega_*} < 2$. The residual potential can therefore be greater than unity. For finite $\bky$ this translates into an effective resonance between scales $\bkx \sim \bky$ (as can be seen by the peak in the left plot in Figure~\ref{fig:residual_compare}). There is therefore an effective resonance between the primary and the zonal response. 


Though this simplified velocity-structure given by Eq.~\eqref{eq:gen_weak_dF_full} captures some of these essential differences from the RH case, it fails short of fully describing the GK residual level quantitatively, as can be seen in Figure~\ref{fig:residual_compare_time_trace}. Agreement can be made closer by taking,
\begin{equation}
    h_{\kx} \approx -i \frac{q}{T} \bky \bkx c
    J_{0, \kx\ky}J_{0,\ky}
    \phi_{\bky}^*\phi_{\bkx}
    \frac{(\hat{\omega}_{d,\ky}-\hat{\omega}_{*,\ky}^T)\omega_{\kx} - (\hat{\omega}_{*,\ky}^T -\omega_{\ky})\hat{\omega}_{d, \kx} }{(\omega_{\ky}^*-\hat{\omega}_{d,\ky})(\omega_{\kx\ky}-\hat{\omega}_{d,\kx\ky})(\omega_{\kx} - \hat{\omega}_{d,\kx})}F_0,
    \label{eq:NL_term_residual}
\end{equation}
ignoring the streaming and $\phi_{\kx}$ terms. 

\begin{figure}
    \centering
    \includegraphics[width=0.8\textwidth]{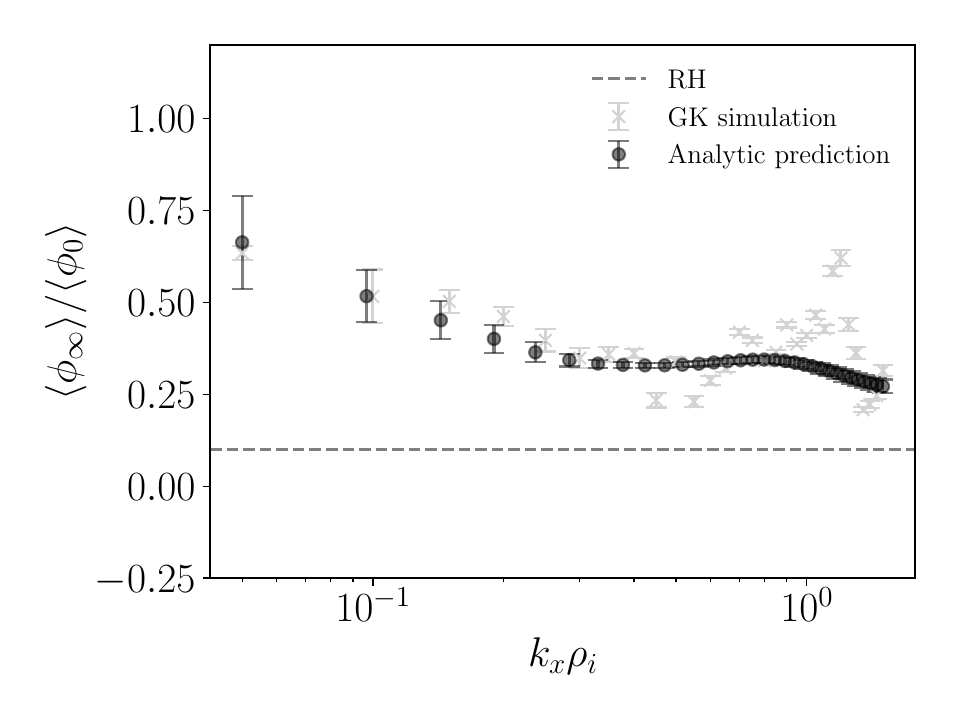}
    \caption{\textbf{Residual Spectrum from the Weakly Nonlinear Phase.} Figure showing the comparison between the residual spectra of the full gyrokinetic simulation (grey) and the analytical prediction (black) given in Eq.~\eqref{eq:NL_term_residual}. Here the vertical bars are the standard deviation in the residual value as calculated over the last one-third of the time domain. Details of the simulation are given in Appendix~\ref{sec:numerical_studies}.}
    \label{fig:Model_for_residual_GK}
\end{figure}

We plot this comparison in Figure~\ref{fig:Model_for_residual_GK}.  Here the $\kx$-spectrum of the residual for the gyrokinetic simulation against the case using Eq.~\eqref{eq:NL_term_residual} as the initial condition for the zonal mode is shown. For the residual of the full gyrokinetic simulation during the $2\gamma$-regime, a full nonlinear gyrokinetic simulation was run, and during the weakly nonlinear regime, the simulation was restarted linearly. 

It is worth noting that this residual spectrum is very sensitive to the precise values and structure of the drifts and drive terms, given an underlying competition between various small quantities. One salient example is the importance of the $z$-structure of Eq.~\eqref{eq:NL_term_residual}. In order to reproduce the comparison in Figure~\ref{fig:Model_for_residual_GK} the $z$-dependence of primary and sideband potentials were required to be fitted from the fully nonlinear GK simulation, and used as inputs for Eq.~\eqref{eq:NL_term_residual}. It is worth noting that by fitting the non-zonal modes numerically, we may be partially capturing some of the streaming effects in the driving modes, but not fully. Numerical resolution can also be playing an important role in reproducing the exact residual spectrum due to this high sensitivity.

The largest discrepancies are seen at high $\kx$. This is likely due to the assumption of using a single frequency for all modes across the $\kx$ spectrum. This is not strictly true, especially at larger $\kx$-values where the sidebands may even be stabilised.  

All of the above raises the question of whether the residual is a meaningful feature in fully developed turbulent state. This also calls into question the utility of the traditionally considered residual.

\section{Conclusion}\label{sec:conclusion}
We have investigated the origin and structure of zonal flows in the weakly nonlinear regime, where the zonal flow is purely driven by unstable linear modes. We introduce an appropriate ordering in amplitudes and time scales in which the so-called $2\gamma$ behaviour is dominant, and consider its minimal description. We compare this directly to nonlinear, electrostatic, gyrokinetic simulations, showing agreement that confirms the validity of its essential features.

Within this framework, it is highlighted that the apparent onset of the zonal flow growth in simulations is an artifact. This distinguishes it from other growth mechanisms, such as secondary instabilities, which we also contextualise on similar terms and orderings. Provided the primary mode is linearly unstable, the $2\gamma$ zonal response will inevitably emerge given sufficient time, as a consequence of nonlinear forcing. Moreover, it will always be present, even if subdominant in amplitude. An expression for the zonal-mode amplitude is derived and shown to be large in magnetic fields with small geodesic curvature.

The phase-space structure of the zonal response is closely tied to that of the driving primary. In the case of ITG drive, this contributes a large temperature perturbation, resulting in a larger zonal flow residual than in the classic Rosenbluth-Hinton scenario. The residual sensitivity to the exact form of the velocity-space structure calls into question its physical relevance. The analysis nevertheless shows that there is a large component of the zonal flow that is not visible through the electrostatic potential, and motivates a broader diagnostic when considering the effects of zonal flows including in the regime of fully developed, saturated turbulence. 

In more complicated geometries, parallel streaming may become important, and thus a natural extension of this work would include its effects, which we have neglected in this paper. Beyond this weak regime, in saturation, it remains to be seen if any remnants of this driven mechanism persists. It is conceivable that the long-wavelength zonal flow would be continuously fed via this mechanism, even when it may not be the dominant behaviour. Future work will aim to address both of these questions. 

\begin{acknowledgments}
\section*{Acknowledgments}
We are indebted to Paul Costello, Linda Podavini, Richard Nies and Per Helander for their helpful discussions on various topics throughout this project.

\section*{Declaration of Interests}
The authors report no conflict of interest.
\end{acknowledgments}
\newpage

\appendix

\section{Residuals and Secondary Instabilities}\label{sec:RDK_appendix}

In this Appendix, we reconsider in the gyrokinetic context of this paper the secondary instability work of Rogers, Dorland, and Kotschenreuther (RDK) \citep{Rogers2000}, and place it alongside the typical Rosenbluth-Hinton (RH) residual response \citep{RosenbluthHinton1998}. We set as our aim the derivation of an evolution equation that can capture both simultaneously, and allows us connections to the new enhanced residual of Section \ref{sec:res_2_gamma}. 

\subsection{Fluid Moment Set-Up}
For the purposes of this Appendix, we reinstate the species indices. We start by introducing $Q_{\kx}$ and $\hq_{\veck}$ such that $h_{\s,\veck } = \hq_{\veck} \exp(-iQ_{\kx})$, and
\begin{equation}
    v_{\parallel}\nabla_{\parallel}Q_{\kx}=k_{\x}\left(\mathbf{v}_{d}\cdot\nabla\x-\bar{v}_{\x}\right)\equiv k_{\x}\left[u^{\x}\left(\frac{\hat{v}_{\perp}^{2}}{2}+\hat{v}_{\parallel}^{2}\right)-\bar{u}^{\x}\hat{\mathcal{E}}\right].,
\label{eq:Q}
\end{equation}
where the overbar here represents a bounce average (which we shall take to vanish, as it must in any omnigeneous field \citep{hall1975three, cary1997omnigenity}. Re-writing the drifts and streaming terms in equation \eqref{eq:GK_real} in terms of $Q_{\kx}$ we get
\begin{equation}
    \frac{\partial}{\partial t}\left(h_{\s}- \frac{q_{\s} F_{0 \s}}{T_{0 \s}} \gyrophi \right) 
    + \vpa \nabla_{\parallel} \left( e^{iQ_{\bkx}} h_{\s}\right) e^{-iQ_{\bkx}}
    =
    - \frac{c}{B} \hat{\vec{b}} \cdot \gyrophi \cross F_{0 \s} 
    - \frac{c}{B} \hat{\vec{b}} \cdot \gyrophi \cross \nabla h_{\s}.
    \label{eq:RDK_gk_h}
\end{equation}
It is important to emphasize that the above equation is evaluated at the gyrocenter position, $\vec{R}$. To proceed we define 
\[
g_{\s \veck} =  h_{\s \veck} -  \frac{q_{\s} F_{0\s}}{T_{0\s}} J_{0\veck} \phi_{\veck}, 
\]
which is also evaluated at gyrocenters. We next define the following guiding centre fluid moments \citep{catto1977linearized, abel2013multiscale}
\begin{equation}
    \frac{\delta \bar{n}_{\veck} \left( \vec{R} \right)}{n_{0\s}} = \frac{1}{n_{0\s}} \int \rmd^3 \vec{v} \Bigg|_{\vec{R}} g_{\s \veck} \qquad  \frac{\delta \bar{T}_{\perp \veck}\left( \vec{R} \right)}{T_{0\s}} = \frac{1}{n_{0\s}} \int \rmd^3 \vec{v} \Bigg|_{\vec{R}} \frac{v_{\perp}^2}{2} g_{\s \veck}
\end{equation}
To write the gyrokinetic equation, Eq.~\eqref{eq:RDK_gk_h}, in terms of fluid moments, we take
\[
\frac{1}{n_{0\s}} \int \rmd^3 \vec{v} \Bigg|_{\vec{R}} e^{iQ_{\kx}} \cdot
\]
and obtain
\begin{align}
        \frac{\partial}{\partial t} \left( \frac{\delta \bar{n}_{\veck}}{n_{0\s}} \right) + &
        \frac{\partial}{\partial t} \frac{1}{n_{0}}\int \rmd^{3}\vec{v} \left(i Q_{\kx} - \frac{Q_{\kx}^2}{2}\right) \vpa g_{\s \veck}
        + \nabla_{\parallel} \int \rmd^3 \vec{v} \left(1+ iQ_{\kx} - \frac{Q_{\kx}^2}{2}  \right) h_{\veck}
        = \nonumber\\
        &- i c \ky \int \rmd^3 \vec{v}\left(1 - \frac{Q_{\kx}^2}{2}  \right) F_{0}^{\prime}J_{0 \veck}\phi_{\veck}  
        \nonumber\\
        &+ c \sum_{\veckp} \int \rmd^3 \vec{v} \left(1+ iQ_{\kx} - \frac{Q_{\kx}^2}{2}  \right) \hat{\vec{b}} \cdot \veck \cross \veckp J_{0, \veck - \veckp} \phi_{\veck - \veckp} g_{\veck^{\prime}} .
    \label{eq:RDK_gk_hk}
\end{align}

\subsection{Recovering RDK}
To make a connection to the RDK equation, Eq.~(1) in \cite{Rogers2000}, let us start by relating the density of guiding centers, $\delta \bar{n}_{\veck}$, to the density at particle positions, $\delta n_{\veck}$, for small $\bbk \ll 1$
\begin{equation}
\begin{split}\frac{\delta n_{\veck}}{n_{0\s}} & =\frac{1}{n_{0\s}}\int\rmd^{3}\vec{v}\Bigg|_{\vec{r}}\left[J_{0\veck}h_{\veck}-\frac{q_{s}\phi_{\veck}}{T_{0\s}}F_{0\s}\right]\\
 & \approx\frac{1}{n_{0\s}}\int\rmd^{3}\vec{v}\left[\left(1-\frac{\bbk\xperp}{2}\right)h_{\veck}-\frac{q_{s}\phi_{\veck}}{T_{0\s}}F_{0\s}\right]\\
 & =\frac{1}{n_{0\s}}\int\rmd^{3}\vec{v}\left[h_{\veck}-\left(1-\frac{\bbk\xperp}{2}\right)\frac{q_{s}\phi_{\veck}}{T_{0\s}}F_{0\s}-\frac{\bbk\xperp}{2}\left(h_{\veck}-\frac{q_{s}\phi_{\veck}}{T_{0\s}}F_{0\s}\right)-\bbk\xperp\frac{q_{s}\phi_{\veck}}{T_{0\s}}F_{0\s}\right]\\
 & =\frac{\delta\bar{n}_{\veck}}{n_{0\s}}-\frac{k_{\perp}^{2}\rho_{\s}^2}{2}\frac{\delta\bar{T}_{\perp\veck}}{T_{0\s}}-\frac{k_{\perp}^{2}\rho_{\s}^2}{2}\frac{q_{s}\phi_{\veck}}{T_{0\s}}\\
\Rightarrow\frac{\delta\bar{n}_{\veck}}{n_{0\s}} & =\frac{\delta n_{\veck}}{n_{0\s}}+\frac{k_{\perp}^{2}\rho_{\s}^{2}}{2}\frac{\delta\bar{T}_{\perp\veck}}{T_{0\s}}+\frac{k_{\perp}^{2}\rho_{\s}^{2}}{2}\frac{q_{s}\phi_{\veck}}{T_{0\s}}.
\end{split}\label{eq:delta_n_GC_r}
\end{equation}
The terms proportional to $k_{\perp}^2\rho_{\s}^2$ manifest the FLR corrections, and are key in the nonlinearity driving the RDK mechanism. However, it is important to note that this relation is only valid in the small, $\bbk \ll 1$, and will eventually break down regardless of how many terms are included in the expansion of the Bessel functions. 

We shall then model the electron response using a modified adiabatic approximation \citep{BillGreg1993, Zocco_Mishchenko_Koenies_Falessi_Zonca_2023}, so that
\begin{equation}
    \frac{\delta \bar{n}_{e}}{n_{0}} = \frac{e}{T_{0e}}\left(\phi-\left\langle \phi\right\rangle _{\psi}\right) + \underbrace{\frac{k_{\perp}^2 \rho_{e}^2}{2} \frac{\delta \bar{T}_{\perp \veck}}{T_{0 e}} + \frac{k_{\perp}^2 \rho_{e}^2}{2} \frac{q_e \phi_{\veck}}{T_{0 e}} }_{\text{small}}. \label{eq:adiab_elect_fluid}
\end{equation}

Using this into Eq.~\eqref{eq:RDK_gk_hk}, neglecting geometry, background gradients, finite orbit width effects, and using the modified adiabatic response approximation for electrons, we then find 
\begin{equation}
    \frac{\rmd}{\rmd t}\left[ \frac{e}{T_{0e}}\left(\phi-\left\langle \phi\right\rangle _{\psi}\right) 
    -\frac{\rho_{i}^{2}\nabla_{\perp}^{2}}{2}\frac{e_i \phi}{T_{0i}}
    -\frac{\rho_{i}^{2}\nabla_{\perp}^{2}}{2}\frac{\delta T_{\perp i}}{T_{0i}}\right]+\frac{1}{2}\left\{ \rho_{i}^{2}\nabla_{\perp}^{2}\phi,\frac{\delta T_{\perp i}}{T_{0i}}\right\} =0, 
    \label{eq:rogersmine}
\end{equation}
where $\{ A, B\} $ indicates the Poisson bracket, defined as $c/B \left(\partial_{\x} A \partial_{\y} B -  \partial_{\x} B \partial_{\y} A \right)$. Using the same notation and normalisations as RDK, and noting that they use electrostatic potential at guiding centers, $\phi \approx (1 - \rho_i^2 \grad_{\perp}^2/ 2) \psi$ which is defined as the electrostatic potential at guiding centers this gives
\begin{equation}
    \frac{\rmd}{\rmd t}\left( \psi-\left\langle \psi\right\rangle -\tau\nabla_{\perp}^{2}\frac{T_{\perp i}}{2}\right) -\frac{\rmd}{\rmd t}\nabla_{\perp}^{2}\left[ (1+ \tau)\psi - \tau \left\langle \psi\right\rangle\right]
    +\frac{1}{2}\left\{ \tau\nabla_{\perp}^{2}\psi,T_{\perp i}\right\} =0,
\end{equation}
which is the cold ion limit $k_{\perp}\rho_i\sim \tau^{1/2}\ll 1$ of their  Eq.~(1). 
which is the limit $\psi-\left\langle \psi\right\rangle \sim k_{\perp}^2\rho_i^2 \ll 1$ of their  Eq.~(1). This is one more approximation that we seem to need to avoid higher order nonlinearities. Another important approximation introduced by the authors is that of considering diamagnetic effects negligible. However, it is perhaps important to remark  that  these effects enter equation~\eqref{eq:guiding_center_continuity} both implicitly, via the nonlinearity of the $Q$-moment of the perturbed distribution function, and  explicitly, through the contributions from the primary distribution function in the nonlinearities and temperature perturbations. These two aspects will be crucial in future studies of turbulent transport.

\subsection{Connection to the ZF Residual}
Going back to the fully fleshed form of Eq.~\eqref{eq:RDK_gk_hk}, it is natural to isolate the surface-averaged ZF response. Taking the flux surface average of that equation just does that. Defining 
\begin{equation}
        \left\langle A\right\rangle _{\psi}=\frac{1}{V^{\prime}}\oint \rmd \alpha\int \frac{\rmd \ell}{B}\sum_{\vec{k}}A_{\vec{k}}e^{i(k_{\psi}\psi+k_{\alpha}\alpha)}
        =\frac{2\pi}{V^{\prime}}\int \rm\frac{\rmd \ell}{B}\sum_{k_{\psi}}A_{k_{\psi}0}e^{ik_{\psi}\psi}, 
\end{equation}
this action annihilates the parallel streaming term, and gives only $\ky = 0$ contributions. The resulting equation is one for the perturbed density of guiding centers. 

Following Dorland and Hammett, we approximate the gyroaverage operators, but in this case keep the finite orbit width (FOW) effect, considering an ordering $Q_{\kx} \sim k_{\perp} \rho_{\s} \ll 1 $. The resulting equation, valid up to order $\sim \mathcal{O}(Q_{k_\x}^2) \sim \mathcal{O}(k_{\perp}^2 \rho_{\s}^2)$,
can be written as follows,
\begin{equation}
    \begin{split}
        \frac{\rmd}{\rmd t} \left\langle \frac{\delta \bar{n}_{\veck} }{n_{0\s}} \right\rangle_{\x} + 
        \frac{\partial}{\partial t} \left\langle \frac{1}{n_{0}}\int \rmd^{3}\vec{v} \left(i Q_{\kx} - \frac{Q_{\kx}^2}{2} \right) g_{\s \veck} \right\rangle_{\x}
        =
        \frac{c}{B} \left\langle
        - \left\{\frac{\rho_{\s}^2 \nabla^2_{\perp}}{2}  \phi, \frac{\delta \bar{T}_{\perp}}{T_{0\s}} \right\}_{\veck} 
        \right. 
        \\
        \left.
        + \left[ \partial_{\x} \left\{ \phi,\frac{1}{n_{0}}\int d^{3}\vec{v}\frac{Q_{\kx}}{\kx}g_{\s \veck} \right\}
        + \partial_{\x}^2 \left\{\phi,\frac{1}{n_{0}}\int d^{3}\vec{v}\frac{Q_{\kx}^2}{\kx^2}g_{\s \veck} \right\} \right]_{\vec{k}}
        \right\rangle_{\x} ,
        \label{eq:guiding_center_continuity}
    \end{split}
\end{equation}
Using the adiabatic electron response, Eq.~\eqref{eq:adiab_elect_fluid}, one obtains
\begin{multline}
        \frac{\rmd}{\rmd t} \left\langle \frac{k_{\perp}^2 \rho_{i}^2}{2} \frac{q_i\phi_{\veck}}{T_{0 i}}  + \frac{k_{\perp}^2 \rho_{i}^2}{2} \frac{\delta \bar{T}_{\perp \veck}}{T_{0 i}} \right\rangle_{\x} + 
        \frac{\partial}{\partial t} \left\langle \frac{1}{n_{0}}\int \rmd^{3}\vec{v} i Q_{\kx} g_{\s \veck} \right\rangle_{\x}     
        \\
        =
        \frac{c}{B} \left\langle
        - \left\{\frac{\rho_{ i}^2}{2} \nabla_{\perp}^2 \phi, \frac{\delta \bar{T}_{\perp}}{n_{0 i}} \right\}_{\veck} 
        + \left[ \partial_{\x} \left\{ \phi,\frac{1}{n_{0}}\int d^{3}\vec{v}\frac{Q_{\kx'}}{\kx'}g_{\s \veck'} \right\}
        \right. \right.
        \\
        \left. \left. 
        + \partial_{\x}^2 \left\{\phi,\frac{1}{n_{0}}\int d^{3}\vec{v}\frac{Q_{\kx'}^2}{\kx'^2}g_{\s \veck'} \right\} 
        \right]_{\vec{k}}
        \right\rangle_{\x} ,
        \label{eq:guiding_center_continuity_sumspec}
\end{multline}
The source term agrees, as it should, with RDK, Eq.~\eqref{eq:rogersmine}, except for the FOW terms. Thus, the equation reconciles the picture of both zonal flows as secondary instabilities with the Rosenbluth-Hinton residual scenario. It is convenient to define the following quantity
\begin{equation}
    W = \left\langle \frac{k_{\perp}^2 \rho_{e}^2}{2} \frac{q_i  \phi_{\veck} }{T_{0 i}}+ \frac{k_{\perp}^2 \rho_{i}^2}{2} \frac{\delta \bar{T}_{\perp \veck}}{T_{0 i}} + \frac{1}{n_{0}}\int \rmd^{3}\mathbf{v}iQ_{k_{\x}}\delta f_{\vec{k}}\right\rangle_{\x},
\end{equation}
which Eq.~\eqref{eq:guiding_center_continuity_sumspec} evolves in time driven exclusively by nonlinearities. When evolved linearly,  $W|_{t=0} = W|_{t\rightarrow \infty}$. Using the pure density form of the distribution function as \cite{RosenbluthHinton1998}, Eq.~\eqref{eq:RH_df_0}, at $t=0$ we have
\begin{equation}
\begin{split}
    W(t=0) & = \Bigg\langle 
    \underbrace{\frac{k_{\perp}^{2}\rho_{i}^{2}}{2}\frac{e\phi_{\veck 0 }}{T_{0i}}}_{\mathcal{O}(b_{\veck})} 
    + \underbrace{\frac{k_{\perp}^2 \rho_{i}^2}{2} \frac{\delta \bar{T}_{\perp \veck0}}{T_{0 i}}}_{\mathcal{O}(b_{\veck}^2)}
    + \underbrace{ \frac{1}{n_{0}}\int \rmd^{3}\mathbf{v}iQ_{k_{\x}}\delta f_{\vec{k} 0} }_{=0} \Bigg\rangle_{\x}
    \\
    & \approx \left\langle \frac{k_{\perp}^{2}\rho_{i}^{2}}{2}\frac{e\phi_{\veck 0 }}{T_{0i}} \right\rangle_{\x},
    \label{eq:W_zero}
\end{split}
\end{equation}
where the final $\df$ integral evaluates to zero because it is off in $v_{\parallel}$. To obtain $W|_{t\rightarrow \infty}$ we need the distribution function in terms of $\phi_{\veck}(t\rightarrow\infty)$. In line with the RH analysis (see Appendix~\ref{sec:ZF_residuals}), we take this to be
\begin{equation}
\begin{split}
    \delta f_{\veck \infty} & = e^{-iQ_{\kx}}\overline{e^{iQ_{\kx}}\df_{0 \veck}} + \frac{qF_0}{T}\left(\overline{J_{0 \veck} e^{iQ_{\kx}}}e^{-iQ_{\kx}}-J_{0 \veck}\right)\phi_{\veck \infty}
    \\
    \Rightarrow \delta f_{\veck \infty} & = \overline{\delta f_{\veck 0}} + \left[
    -i (Q_{\kx}-\overline{Q_{\kx}}) + \overline{Q_{\kx}}Q_{\kx} - \frac{1}{2} \left(Q_{\kx}^2 + \overline{Q_{\kx}^2} \right) + \frac{k_{\perp}^2 \rho_i^2 - \overline{k_{\perp}^2 \rho_i^2}}{4}\hat{v}_{\perp}^2
    \right] \frac{e \phi_{\veck \infty}}{T_{0 i}} F_{0 i}, 
    \label{eq:deltaf_inft}
\end{split}
\end{equation}
where $\overline{(\cdot)}= \left( \int \rmd \ell \sqrt{g}\cdot/v_{\parallel} \right) / \left(\int \rmd \ell \sqrt{g}/v_{\parallel}\right)$ denotes the bounce average. Using this, we find
\begin{equation}
    W(t\rightarrow \infty) \approx \Bigg\langle 
    \underbrace{\frac{k_{\perp}^{2}\rho_{i}^{2}}{2}\frac{e\phi_{\veck \infty }}{T_{0i}}}_{\mathcal{O}(b_{\veck})}
    + 
    \underbrace{\frac{k_{\perp}^2 \rho_{i}^2}{2} \frac{\delta \bar{T}_{\perp \veck \infty}}{T_{0 i}}}_{\mathcal{O}(b_{\veck}^2)}
    + 
    \underbrace{\frac{1}{n_0} \int \rmd^3 v Q_{\kx}(Q_{\kx} - \overline{Q}_{\kx}) F_{0i} \frac{e\phi_{\veck \infty}}{T_{0i}}}_{\mathcal{O}(b_{\veck})}
    \Bigg\rangle_{\x}.
    \label{eq:W_infinity}
\end{equation}
Equating the initial, Eq.~\eqref{eq:W_zero}, and final, Eq.~\eqref{eq:W_infinity}, values of $W$, we obtain an expression for the residual level normalised to the initial amplitude
\begin{equation}
    \frac{\phi_{0}(t\rightarrow\infty)}{\phi_{0}(t=0)} = \frac{\left\langle \frac{k_{\perp}^{2}\rho_{i}^{2}}{2} \right\rangle_{\x} }{\left\langle \frac{k_{\perp}^{2}\rho_{i}^{2}}{2}
    + \frac{1}{n_0} \int \rmd^3 v Q(Q - \overline{Q}) F_{0i}\right\rangle_{\x} },
    \label{eq:residual_level_RDK_ap}
\end{equation}
which is the standard form of the residual in an omnigeneous stellarator \citep{RosenbluthHinton1998, mishchenko2008collisionless, monreal2016residual,plunk2024residual}.

A similar analysis can be carried out by applying the same procedure to the modified initial condition derived in Section~\ref{sec:res_2_gamma}. In this case, the ordering of the temperature contribution changes compared to Eq.~\eqref{eq:W_zero}
\begin{equation}
    W(t=0) = \Bigg\langle     \underbrace{\frac{k_{\perp}^{2}\rho_{i}^{2}}{2}\frac{e\phi_{\veck 0 }}{T_{0i}}}_{\mathcal{O}(b_{\veck})}
    + \underbrace{\frac{k_{\perp}^2 \rho_{i}^2}{2} \frac{\delta \bar{T}_{\perp \veck0}}{T_{0 i}}}_{\mathcal{O}(b_{\veck})}
    - \underbrace{ \frac{1}{n_{0}}\int \rmd^{3}\mathbf{v}iQ_{k_{\x}}\delta f_{\vec{k} 0} }_{=0} \Bigg\rangle_{\x}.
    \label{eq:W_zero2}
\end{equation}
Following this, it is conceivable that a larger initial temperature perturbation leads to a larger residual level. However, the initial $\delta T_{\perp}$ perturbation cannot change by more than $\mathcal{O}(b_{\veck})$ following Eq.~\eqref{eq:deltaf_inft}. Given the conservation of $W$, this means that the change to $\phi$ cannot be more than $\delta \phi_{\veck} /\phi_{\veck 0} \sim \delta T_{\perp\veck 0} b_{\veck} /(e\phi_{\veck 0}) $ as a result. Therefore, to see the effects of the $\delta T_{\perp \veck 0}$ perturbation one would need $e\phi_{\veck 0 }/T_{0,i} \sim \mathcal{O}(b_{\veck})$, which is a different ordering to the one considered above, and would require going to next order.





\section{Nies' Secondary Mode}\label{sec:toroidal_secondary_appendix}

\begin{figure}
    \centering
    \includegraphics[width=0.45\linewidth,keepaspectratio]{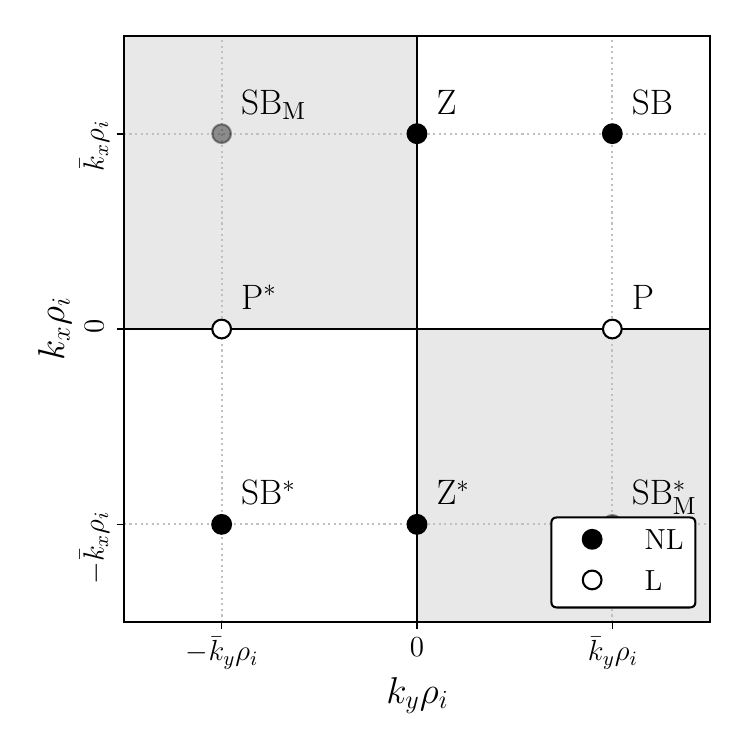}
    \caption{\textbf{Minimal Model for Strongly Nonlinear Regime}. $\vec{k}$-space diagram illustrating the modes used to describe the ZF dynamics in the strongly nonlinear regime.}
    \label{fig:strongly_nl_minimal_mode}
\end{figure}

In this Appendix, the work of \cite{Richard2024} is cast in the language of the minimal-model framework used in this paper. Their work falls under the strongly nonlinear regime, see Section~\ref{sec:Regimes}, for which we require one mode in addition to the three used in the main text (see Figure~\ref{fig:strongly_nl_minimal_mode} compared to Figure~\ref{fig:compare_regimes}). This additional mode corresponds to the mirror counterpart of the sideband.

The corresponding set of governing equations are Eqs.~\eqref{eq:schematic_GK} with nonlinear terms as given by
\begin{align}
	\begin{cases}
        \displaystyle
		\mathcal{N}_{k_y} & = 0, \\
        \displaystyle
        \mathcal{N}_{k_x k_y} & = ic k_x k_y
		\left[ J_{0,k_x}\,\phi_{k_x}\,\dgg_{k_y} - J_{0,k_y}\,\phi_{k_y}\,\dgg_{k_x} \right], \\
        \displaystyle
        \mathcal{N}_{k_x, -k_y} & = \mathcal{N}_{k_x k_y}^*, \\
		\displaystyle
		\mathcal{N}_{k_x} & = ic k_x k_y \left[J_{0,k_y}\,\phi_{-k_y}\,\dgg_{k_x k_y}- J_{0,k_x k_y}\,\phi_{k_x k_y}\,\dgg_{-k_y} \right] ,
	\end{cases}
\end{align}
where we have adopted the notation used by \cite{Richard2024}\footnote{Here, $\{x, y\}$ are the usual normalised code variables corresponding to $\{\psi, \alpha\}$ \citep[Chapter 4]{Richard2024}.}. 
Let us consider the equation for the ZF explicitly, using $g\stackrel{\cdot}{=}\langle\delta f\rangle_\mathbf{R}=h-(q\phi/T)J_0F_0$ in line with the work of \cite{Nies2025Thesis},
\begin{multline}
    \partial_t g_{k_x}+i k_x v_{d, x}\left(g_{k_x}+\frac{q}{T}J_{0 k_x}F_0\phi_{k_x}\right) +ik_y v_{E,y}^{k_x, -k_y}J_{0k_x, -k_y}g_{k_y}\nonumber
    \\   
    +ik_x v_{E,x}^{k_y}J_{0 k_y}g_{k_x, -k_y}-ik_y v_{E,y}^{k_x k_y}J_{0 k_x} g_{ -k_y}+ik_x v_{E,x}^{ -k_y}J_{0 k_y}g_{k_x k_y}=0 ,
\end{multline}
with
\begin{subequations}
    \begin{align}
        v_{E,x}^{k_y}&=-ik_y \frac{\mathbf{b}\times\nabla x\cdot\nabla y}{B}\phi_{k_y}, \\
        v_{E,y}^{k_x}&=-ik_x \frac{\mathbf{b}\times\nabla x\cdot\nabla y}{B}\phi_{k_x} .
    \end{align}
\end{subequations}
In the limit $k_x \rho_i\gg k_y \rho_i$, the sideband and zonal modes are strongly coupled, and we can approximate $\phi_{k_x k_y} \approx \phi_{k_x}$, $\phi_{k_x -k_y} \approx \phi_{k_x}$. With this simplification, and dropping any finite orbit width effects associated to $k_y$,
\begin{multline}
    \partial_t g_{k_x}+ik_x v_{d, x}\left(g_{k_x}+\frac{q}{T}J_{0 k_x}F_0\phi_{k_x}\right)+ik_x (v_{E,x}^{k_y} 
    +v_{E,x}^{ -k_y})g_{k_x}
    \\
    + ik_y v_{E,y}^{k_x}J_{0 k_x}(g_{k_y}-g_{ -k_y})\approx 0.
\end{multline}
Using $g_{ -k_y}=g_{k_y}^*$ and $\phi_{ -k_y}=\phi_{k_y}^*$, and further taking $\partial_t\rightarrow-i\omega$, we obtain
\begin{equation}
    g_{k_x}=\frac{k_x v_{d, x}-2k_x \mathrm{Re}[v_g^{k_y}]}{\omega -k_x v_{d, x}-2k_x \mathrm{Re}[v_{E,x}^{k_y}]}J_{0 k_x}\phi_{k_x}\frac{qF_0}{T},
\end{equation}
where we have introduced the normalised form for the derivative of the distribution function
\begin{equation}
    v_g^{k_y}=-\frac{k_y }{k_x }\frac{g_{k_y}/F_0}{q\phi_{k_x}/T}v_{E,y}^{k_x} .
\end{equation}
This is precisely the same form of the distribution function associated to the secondary zonal response as in \cite[Eq.~(27)]{nies2026theory}, with the only difference that there the authors treated $v_E$ and $v_g$ as functions of $y$, while here they are real scalar quantities. This difference disappears if one truncates their functions of $y$ and take the limit $k_y \rightarrow 0$. Although this may seem a crude approximation (enforcing perfect coherence between $(k_x, k_y)$, $(k_x, -k_y)$ and $(k_x,0)$), it yields precisely the same results, as the resulting equations are formally equivalent. Given this equivalence, the dispersion relations of the toroidal secondary and RDK in \cite{Nies2025Thesis, Richard2024} follows from taking quasineutrality. 

\section{Laplace Derivation} \label{sec:laplace_derivaion}
In this Appendix, we analyse the time behaviour in the weakly nonlinear regime using a Laplace analysis. This is done in analogy with Section~\ref{sec:weaklyNL} to carefully account for the time response of the system and consider subdominant behaviour, as well as the linear time response of the zonal mode.

\begin{figure}
    \hspace{3.2cm}
    \includegraphics[width=0.7\linewidth]{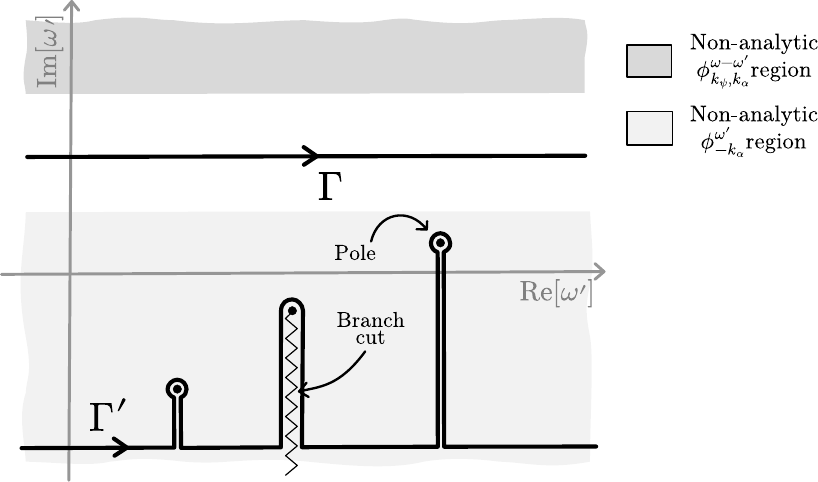}
    \caption{Contour $\Upgamma$ in the complex $\omega^{\prime}$ plane. The imaginary part of $\omega$ is taken to be sufficiently positive so that there exists a region in which both $\phi_{\kx \ky}^{\omega - \omegap}$ and $\phi_{-\ky}^{\omegap}$ are analytic and well behaved. The contour $\Upgamma$ is chosen within this region. It may then be deformed toward $\operatorname{Im}[\omega^{\prime}] \to -\infty$, contour $\Upgamma'$, at which point all poles of $\phi_{-\ky}^{\omegap}$ (in the light gray region) are enclosed and their contributions are picked up.}
    \label{fig:contour}
\end{figure}

We begin by considering the Laplace transform of the ZF GK equation, Eq.~\eqref{eq:zonal}. For this, we introduce the Laplace transform, 
$\phi^{\omega}=\mathcal{L}[\phi(t)]=\int_{0}^{\infty} \rmd t \, \phi(t)\exp(i\omega t),$ with $\Im[\omega]>\sigma, \sigma\in\mathbb{R},$ and $\exp(-\sigma t)\left|\phi(t)\right|<M$, for all $t$ and sufficiently large $M\in\mathbb{R}$. We set aside the response from initial conditions, which evolves purely linearly as its response is decoupled from the nonlinear drive in the weakly nonlinear regime. This linear response is clearly seen at the beginning of the simulations in Figure~\ref{fig:different_initial_conditions}, leaving the weakly nonlinear behaviour unaffected. The resulting equation is then
\begin{equation}
    (\omega - \hatomegadx) \, \dgg_{\bkx}^{\omega} = \omega \frac{q \maxwellian}{T_{0}} \besselk \phi^{\omega}_{\bkx} - \bar{\mathcal{N}} \int_{\Upgamma} \frac{\rmd \omegap}{2\pi i}  \left[ J_{0,\bky}\,\phi_{-\bky}^{\omegap} \,\dgg_{\bkx \bky}^{\omega - \omegap} - J_{0,\bkx \bky}\,\phi_{\bkx \bky}^{\omega - \omegap}\,\dgg_{-\bky}^{\omega} \right], 
\end{equation}
where the integral over $\omegap$ arises from the convolution in frequency space and must be taken along $\Upgamma$, an integration path that must be chosen within the shared region of analyticity of the convolved functions (see Figure~\ref{fig:contour}). Provided a sufficiently large $\sigma$ this region is guaranteed to exist.

Following the definition of the weak regime, we take the non-zonal components to follow their respective linear GK equations, Eq.~\eqref{eq:nonzonal}, and neglect the parallel streaming contribution. From this we obtain the following expression:
\begin{multline}
    (\omega - \hatomegadx) \, \dgg_{\bkx}^{\omega} =\omega \frac{q \maxwellian}{T_{0}} \besselk \phi_{\bkx}^{\omega} \\
    - \bar{\mathcal{N}} \int_\Upgamma \frac{\rmd \omega'}{2\pi i}  J_{0,\bky} J_{0,\bkx \bky} \left[ \frac{\mathcal{R}_{\bkx \bky}(\omega - \omegap)}{\mathcal{L}_{\bkx \bky}(\omega - \omegap)} - \frac{\mathcal{R}_{-\bky}(\omegap)}{\mathcal{L}_{-\bky}(\omegap)}\right] \frac{q \maxwellian}{T_{0}} \phi_{-\bky}^{\omegap} \phi_{\bkx \bky}^{\omega- \omegap}.
\end{multline}
Applying quasineutrality yields an analogous expression to Eq.~\eqref{eq:zonal_qn_deltafn}, but recast in the Laplace formalism
\begin{equation}
    \epsilon_{\bkx}(\omega) \phi_{\bkx}^{\omega} = i c \bkx\bky \int \rmd^{3}\mathbf{v} \left( \frac{F_{0i}}{n_{0}}
    \frac{J_{0, \bkx} J_{0, \bky}J_{0, \bkx\bky}}{\omega-\hatomegadx} \mathcal{I}_{\omega} \right), 
    \label{eq:laplace_eps}
\end{equation}
with 
\begin{subequations}
\begin{align}
    &\epsilon_{\bkx}(\omega) = \int \rmd^{3}\mathbf{v}\frac{F_{0i}}{n_{0}}J_{0, \bkx}^{2} \frac{\omega}{\omega-\hatomegadx}-1 ,
    \label{eq:zonal_qn_laplace_eps}
    \\
    &\mathcal{I}_{\omega} = \int_{\Upgamma} \frac{\rmd \omegap}{2\pi i} \Bigg[ \Bigg(\underbrace{
        \frac{\omega - \omegap-\hat{\omega}_{*, \bky}}{\omega - \omegap-\hatomegadxy} 
        }_{\textcircled{1}}
        -
        \underbrace{
        \frac{\omegap-\hat{\omega}_{*,-\bky}}{\omegap-\hatomegadmy}
        }_{\textcircled{2}}
    \Bigg)
    \phi_{\bkx\bky}^{\omega-\omegap} \phi_{-\bky}^{\omegap}
    \Bigg],
    \label{eq:zonal_qn_laplace_I}
    \end{align}
\end{subequations}
where $\hat{\omega}_{*, \ky} = \omega_{*, \ky} \left[ 1 + \eta (\hat{v}^2 - 3/2) \right]$, $\eta = \rmd(\ln T) / \rmd (\ln n)$, and $\hat{\omega}_{d, \lambda} = \omega_{d, \lambda} \left(\hat{v}_{\parallel}^{2}+ \hat{v}_{\perp}^{2}/2 \right)$, and $\hat{v} = v / \vths $, as before.
The main difference between this approach and that in Section~\ref{sec:weaklyNL} is the form of the non-zonal potentials. In the Fourier approach they were taken to be $\delta$-functions, whereas here they have a broad time response in frequency space. 

The integral in $\epsilon_{\kx}$ gauges the linear response of the ZF, and it is of precisely the same form as that in the main text, Eq.~\eqref{eq:epsilon_fourier}. The nonlinear integral is, however, more complicated. To evaluate it, we exchange the order of integration, which is permissible under the assumption that the integral is absolutely convergent, noting that the potentials are velocity independent.

Performing a partial fraction decomposition to isolate terms such that their velocity-dependent part in the denominator takes the form $\omega - \hat{\omega}_d$, 


\begin{equation}
    \mathcal{F}(\omega) = \int_{\Upgamma} \frac{\rmd \omegap}{2\pi i} \phi_{\bkx\bky}^{\omega-\omegap} \phi_{-\bky}^{\omegap} \frac{\omegadx}{\omegadx \omegap + \omega \omegady} \int \rmd^{3}\mathbf{v} \frac{F_{0i}}{n_{0}}\overline{J}_0\left( A + B 
    \right),
\end{equation}
where $\overline{J}_0 = \besseltrio$ and
\begin{subequations}
    \begin{align}
        A =&  \frac{\omega - \hatomegastary}{\omega - \hatomegadx} 
    - \frac{\omegap}{\omega}\frac{\omega}{\omega - \hatomegadx} 
    - \frac{\omegadxy}{\omegadx} \frac{(\omega - \omegap) - \hatomegastary}{(\omega - \omegap) - \hatomegadxy},
    \\
    B = & 
    \frac{\omega - \hatomegastarmy}{\omega - \hatomegadx} 
    + \frac{\omegap - \omega}{\omega} \frac{\omega}{\omega - \hatomegadx} 
    - \frac{\omegadmy}{\omegadx} \frac{\omegap - \hatomegastarmy}{\omegap - \hatomegadmy} 
    .
    \end{align}
\end{subequations}
Using the definition of $I_{\vec{k}}^\omega$ in Eq.~\eqref{eq:integral_definitions} (see also Appendix \ref{sec:resonant_nl_integral}), we rewrite this explicitly as 
\begin{subequations}
\begin{align}    
    \int \rmd^{3}\mathbf{v} \frac{F_{0i}}{n_{0}} \overline{J}_0 A  = I_{\bkx}^{\omega}\left(\omegastary \right) 
    - \frac{\omegap}{\omega} I_{\bkx}^{\omega}
    - \frac{\omegadxy}{\omegadx} I_{\bkx \bky}^{\omega - \omegap},
    \\ 
    \int \rmd^{3}\mathbf{v} \frac{F_{0i}}{n_{0}} \overline{J}_0 B 
    =
    I_{\bkx}^{\omega} \left(\omegastarmy \right) 
    + \frac{(\omegap - \omega)}{\omega} I_{\bkx}^{\omega}
    - \frac{\omegadmy}{\omegadx} I_{-\bky}^{\omegap}.
\end{align}
\end{subequations}
Combining $A + B$, and recognising $\omegadx I_{\bkx}^{\omega}\left(\omegastary \right) + \omegadx I_{\bkx}^{\omega} \left(\omegastarmy \right) = 2\omegadx I_{\bkx}^{\omega}$, we get
\begin{align}
    \mathcal{F}(\omega) = &  \int_{\Upgamma} \frac{\rmd \omegap}{2\pi i} \frac{\phi_{\bkx\bky}^{\omega-\omegap} \phi_{-\bky}^{\omegap}}{\omegadx \omegap + \omega \omegady}\left[
    - \omegadxy I_{\bkx \bky}^{\omega - \omegap} 
    + \omegadx I_{\bkx}^{\omega}
    - \omegadmy I_{-\bky}^{\omegap}
    \right]
\\
\equiv & \int_{\Upgamma} \frac{\rmd \omegap}{2\pi i}  \frac{\phi_{\bkx\bky}^{\omega-\omegap} \phi_{-\bky}^{\omegap}}{\omegadx \omegap + \omega \omegady}
\mathcal{M}(\omega, \omegap) .
\end{align}
To evaluate the integral along $\Upgamma$, it is convenient to deform the contour into $\Upgamma'$ by taking it to the region in $\omega'$ space with $\operatorname{Im}[\omega^{\prime}] \to -\infty$ (see Figure \ref{fig:contour}). Note that the contour could have been deformed in the opposite direction, but this would leave the final result unchanged. 

As we deform the contour downwards, we collect the contributions from the enclosed poles and branch cuts (see Figure~\ref{fig:contour}). All other contributions to the integral vanish, as the integrand goes like $\phi_{-\bky}^{\omegap} \phi_{\bkx \bky}^{\omega - \omegap} I_{\pm \bky}(\omegap) \sim 1/\omega^{\prime 3}$ for $|\omegap| \gg 1$\footnote{For $|\omegap| \gg 1$ we have $\phi_{\bky} \sim 1/\omegap$, $\phi_{\bkx \bky} \sim 1/\omegap$, from their linear responses, and $I_{\pm \bky}(\omegap) \sim 1/\omegap$, 
so that $\phi_{\bky}^{\omegap} \phi_{\bkx \bky}^{\omega - \omegap} I_{\pm \bky}(\omegap) \sim 1/\omega^{\prime 3}$.}. The contributions are then: (i) the poles from $\phi_{-\bky}^{\omegap}$, (ii) the zeros of $\omegadx \omegap + \omega \omegady$, and (iii) the branch cuts of $\mathcal{M}(\omega,\omegap)$. Explicitly evaluating the contribution from the simple poles applying Cauchy's Residue Theorem, and symbolically writing the contributions from branchcuts as $\Upgamma_{BC}$,

\begin{equation}
    \mathcal{F}(\omega) = \underbrace{\frac{\phi_{\bkx\bky}^{\omega + \omega_{\bky}^*} \hat{\phi}_{-\bky}^{-\omega^*_{\bky}} }{ \omega \omegady - \omegadx \omega^*_{\bky} }\mathcal{M}(\omega, - \omega_{\bky}^*)}_{\text{(i)}} + 
    \underbrace{\frac{1}{\omegadx}\phi_{\bkx\bky}^{\omega + \omega_1} \phi_{-\bky}^{-\omega_1} \mathcal{M}\left( \omega, - \omega_1 \right)}_{\text{(ii)}} + \Upgamma_{BC}, 
\end{equation}
where $\omega_1 = \frac{\omegady}{\omegadx}\omega$ comes from (ii), and $\hat{\phi}_{\veck}^{\omega}$ indicates the residual value of the function $\phi_{\veck}^{\omega}$. We can then rewrite Eq.~\eqref{eq:laplace_eps} as
\begin{equation}
    \phi_{\bkx}^{\omega} = i c \bkx\bky \frac{\mathcal{F}(\omega)}{\epsilon_{\bkx}(\omega)}.
    \label{eq:ampl_phi}
\end{equation}
To reconstruct the time evolution of $\phi_{\bkx}^{\omega}$ one would inverse Laplace transform Eq.~\eqref{eq:ampl_phi}. We interpret the poles of this expression as pseudo-modes of the ZF, and summarise them in Table~\ref{tab:poles}, anticipating branch cut contributions to be subdominant. 

\begin{table}
\begin{center}
\setlength{\tabcolsep}{14pt}
\begin{tabular}{c | c}
        $\omega = \omega_{\bkx}$ 
         
        & $\mbox{Zero of} \quad \epsilon_{\bkx}(\omega)$
        \\ 
        $\omega = \omega_{\bkx \bky} - \omega_{\bky}^*$ 
         
        & $\mbox{Pole of} \quad \phi_{\bkx\bky}^{\omega + \omega_{\bky}^*}$
        \\ 
        $\omega = \frac{\omegadx}{\omegady} \omega_{\bky}^*$ 
         
        & $\mbox{Zero of } (\omega \omegady - \omegadx \omega^*_{\bky}) \mbox{ and Pole of } \phi_{-\bky}^{-\omega_1}$
        \\ 
        $\omega = \frac{\omegadx}{\omegadxy} \omega_{\bkx \bky}$ 
         
        & $\mbox{Pole of} \quad \phi_{\bkx\bky}^{\omega + \omega_1}$
        \\ 
        $\omega = 0$ 
         
        & $\mbox{Pole of} \quad \mathcal{M}\left( \omega, - \omega_1 \right)$
\end{tabular}
\end{center}
\caption{\textbf{Table of pseudo-modes from Laplace analysis.} The table shows the complex zonal frequencies found from performing a Laplace analysis (left column), and a brief description of their origin from Eq.~\eqref{eq:ampl_phi} (right column). }
\label{tab:poles}
\end{table}

For a typical tokamak geometry $\omegadx \ll \omegady$ in the region where the non-zonal modes are unstable, making $\omega = \omega_{\bkx \bky} - \omega_{\bky}^*$ the root with the largest positive growth. This mode is always present provided the non-zonal modes are unstable (which is assumed in the weakly nonlinear regime). This emphasises the lack of a marginal stability point. In this scenario, all other roots $\sim \omega_{\bky}$ or less. However, it is conceivable that for larger $\kx \rho_i$, other non-zonal linear instabilities, or different geometry $\omega = \omega_{\bkx \bky} - \omega_{\bky}^*$ is no longer the dominant time response. In so far this has not been observed, but will be verified in future studies.

The amplitudes associated with each of these \enquote{modes} can be evaluated from the residual of each of the poles in the inverse Laplace transform of Eq.~\eqref{eq:ampl_phi}. Applied to the dominant one, this yields an expression equivalent to that obtained through the simpler Fourier approach in Section~\ref{sec:weaklyNL}, Eq.~\eqref{eq:zonal_qn_deltafn}. 

\section{Resonant Nonlinear Integral}\label{sec:resonant_nl_integral}

In this Appendix, we re-express the integrals in $ \Omega_{NL}^{-1}$ in Eq.~\eqref{eq:NLresonance}.
The staring integral is repeated here for a drift frequency $\hat{\omega}_\lambda$,
\begin{equation}
    \mathcal{J_{\lambda}} = \int \rmd^{3}\mathbf{v}\frac{F_{0i}}{n_{0}}\overline{J}_0 \frac{1}{\omega-\hat{\omega}_{\psi}}\frac{\omega-\hat{\omega}_{*, \beta}^{T}}{\omega-\hat{\omega}_{\lambda}} 
\end{equation}
where $\overline{J}_0 = \besseltrio$. We have also taken $\eta\gg 1$, such that $\hat{\omega}_{*, \beta} = \omega_{*, \beta}^T (\hat{v}^2 - 3/2)$. Separating the integrand into partial fractions
\begin{equation}
    \begin{split} 
    \mathcal{J_{\lambda}}
    = & \frac{1}{\omega_{\lambda}}\int \rmd^{3}\mathbf{v}\frac{F_{0i}}{n_{0}} \overline{J}_0 \frac{ \omega/\omega_{\psi}-\hat{\omega}_{*, \beta}^{T}/\omega_{\psi}}{\omega/\omega_{\psi}-\left(\hat{v}_{\parallel}^{2}+\frac{\hat{v}_{\perp}^{2}}{2}\right)}\frac{1}{\omega/\omega_{\lambda}-\left(\hat{v}_{\parallel}^{2}+\frac{\hat{v}_{\perp}^{2}}{2}\right)} 
    \\
    = & \frac{1}{\omega_{\lambda}}\int \rmd^{3}\mathbf{v}\frac{F_{0i}}{n_{0}}\frac{\omega_{\psi}-\omega_{*, \beta}^{T}}{\omega_{\lambda}-\omega_{\psi}}\left\{ \frac{\overline{J}_0}{\omega_{\psi}/\omega -\left(\hat{v}_{\parallel}^{2}+\frac{\hat{v}_{\perp}^{2}}{2}\right)}-\frac{\overline{J}_0}{\omega_{\lambda}/\omega -\left(\hat{v}_{\parallel}^{2}+\frac{\hat{v}_{\perp}^{2}}{2}\right)}\right\} 
    \\
    = & \frac{\omega_{\psi}}{\omega\left(\omega_{\psi}-\omega_{\lambda}\right)}\left\{ \int \rmd^{3}\mathbf{v}\frac{F_{0i}}{n_{0}}\overline{J}_0 \frac{\omega-\hat{\omega}_{*, \beta}^{T}}{\omega-\hat{\omega}_{\psi}} -\frac{\omega_{\lambda}}{\omega_{\psi}}\int \rmd^{3}\mathbf{v}\frac{F_{0i}}{n_{0}}\overline{J}_0 \frac{\omega-\hat{\omega}_{*, \beta}^{T}}{\omega-\hat{\omega}_{\lambda}}  \right\} 
    \\ 
    \equiv &
    \frac{\omega_{\psi}}{\omega\left(\omega_{\psi}-\omega_{\lambda}\right)}\left\{ 
    I_{\kx}^{\omega}\left( \omega_{*, \beta}^T \right)
    -\frac{\omega_{\lambda}}{\omega_{\psi}}I_{\lambda}^{\omega} \left(\omega_{*, \beta}^T\right)
    \right\},
    \end{split}
\end{equation}
where
\begin{equation}
    I_{\lambda}^{\omega} \left( \omega_{*, \beta}^T \right) = \int \rmd^{3}\mathbf{v} \frac{F_{0i}}{n_{0}} \overline{J}_0 \frac{\omega - \omega_{*, \beta}^T}{\omega - \hat{\omega}_{d,\lambda}} .
\end{equation}
When the argument $\omega_{*, \beta}^T$ is omitted, then it is understood that $\beta \equiv \lambda$. When finite Larmor radius effects are dropped, $\besselk \approx 1$, the integral can be written as \citep{Biglari89,zocco2018,ivanov2023analytical}
\begin{equation}
    \begin{split} & I_{\veck}^{\omega} \left( \omega_{*,\ky} \right) 
    =\left(1-\frac{\omega_{*,\ky}}{\omega}\right)\frac{\omega}{\omega_{d,\veck}}Z^{2}\left(\sqrt{\frac{\omega}{\omega_{d, \veck}}}\right)\\
    & +\eta_{i}\frac{\omega_{*,\ky}}{\omega}\left\{ \left(1-2\frac{\omega}{\omega_{d, \veck}}\right)\frac{\omega}{\omega_{d, \veck}}Z^{2}\left(\sqrt{\frac{\omega}{\omega_{d, \veck}}}\right)-2\left(\frac{\omega}{\omega_{d, \veck}}\right)^{3/2}Z\left(\sqrt{\frac{\omega}{\omega_{d, \veck}}}\right)\right\}.
\end{split}
\end{equation}

\section{Comparison to \cite{ChenQiuZonca2024}}\label{sec:comparison_to_previous_wrk}
In this Appendix, we briefly compare our work to that of \cite{ChenQiuZonca2024}. In \cite{ChenQiuZonca2024}, they combine the driven (beat-driven) and modulational instability mechanisms to describe the generation of ZFs. 

In the same way that we do in this paper, the authors rely on a truncated mode description. They decompose the perturbed electrostatic potential into a non-zonal (drift-wave) component and a zonal component, with zonal modes defined by vanishing parallel and binormal wavenumbers. This equates to a small set of interacting modes, consisting of a primary drift wave, its radial sidebands (including the mirror mode discussed in Appendix~\ref{sec:toroidal_secondary_appendix}), and a zonal mode.

Their \enquote{primary} treatment differs in that they retain a zero-frequency drift wave component. This leads to fundamental differences in the nonlinearities responsible for driving the zonal mode. In \cite{ChenQiuZonca2024}, the zonal flow is driven by the interaction between a constant component of the drift wave and a component growing at the zonal growth rate, as discussed in the analysis preceding their Eq.~(22). A $2\gamma$-driven interaction would exist within their model, however this occurs at $\kx$ that lies outside their truncation, and is thus ignored. 

A second point of difference is in the time scale ordering assumed. In deriving their \cite[Eq.~15]{ChenQiuZonca2024} the authors assume $\omega_r \gg \gamma$ for the primary drift wave, whereas the analysis presented in this paper takes $\omega_r \sim \gamma$. This leads to some differences in the region of validity and the resulting ZF behaviour.  

\section{Concerning the Zonal Flow residual}\label{sec:ZF_residuals}
In this Appendix we give some of the details necessary to derive the expressions and conclusions reached in Section~\ref{sec:res_2_gamma} in the paper.

\subsection{General form of the residual}
We first, for completeness, present the calculation for the general form of the zonal flow residual. Different forms of this derivation can be found in multiple pieces of work (see e.g., \cite{RosenbluthHinton1998, XiaoCatto2006, SugamaWatanabe2006}), here we proceed with as general as possible an approach. We omit the $\veck$ and species indices for brevity, but the $\veck$-indices shall be reinstated later. Using the linearised form of the gyrokinetic equation as it applies to the zonal mode (\textit{i.e.}, $k_{\y} =0$), and defining $v_\parallel \nabla_\parallel Q_{\kx}=\omega_d$,

\begin{equation}
    \partial_t\left(h-\frac{qF_0}{T}J_{0}\phi\right)= - v_\parallel\nabla_\parallel\left(e^{iQ_{\kx}}h\right)e^{-iQ_{\kx} }, \label{eqn:GK_res}
\end{equation}
assuming omnigeneity, $\overline{\omega}_{d, \kx}=0$, where the overbar denotes a bounce average. At $t\rightarrow\infty$ a steady state requires vanishing of the right-hand-side of Eq.~(\ref{eqn:GK_res}), which means that $h_\infty\stackrel{\cdot}{=}h_{t\rightarrow\infty}=\hq_\infty e^{-iQ_{\kx}}$ where $\hq_\infty=\overline{\hq}_\infty$. With this at hand, we may then take Eq.~(\ref{eqn:GK_res}), bounce average it, and integrate it in time from $t=0$ to $\infty$, to yield
\begin{equation}
    h_{\infty}=\frac{q F_0}{T}\overline{J_{0} e^{iQ_{\kx}}\phi_{\infty}}e^{-iQ_{\kx}}+e^{-iQ_{\kx}}\overline{\langle \df_{0}\rangle_\mathbf{R}e^{iQ_{\kx}}},
\end{equation}
where we have used $\langle\df\rangle_\mathbf{R}=h-(q\phi/T)J_0F_0$, and the subscript 0 in $\df$ represents the initial perturbation of the distribution function at $t=0$.  

We are then in a position to apply quasineutrality to $g_\infty$. Using the modified adiabatic response in which $\delta n_e/n_e=(q_e/T_e)(\phi-\langle\phi\rangle_\psi)$, and taking the flux surface average
\begin{equation}
    \frac{q\langle \phi_{\infty}\rangle_\psi}{T}=\frac{1}{n_{0}}\left\langle\int\mathrm{d}^3\mathbf{v}J_{0} h_\infty\right\rangle_\psi,
\end{equation}
can be rearranged into the form of the residual,
\begin{equation}
    \frac{q\langle \phi_{\infty}\rangle_\psi}{T}=\frac{\frac{1}{n_{0}}\left\langle \int\mathrm{d}^3\mathbf{v}J_{0}e^{-iQ_{\kx}}\overline{e^{iQ_{\kx}}\langle\delta\!f_0\rangle_\mathbf{R}} \right\rangle_\psi}{1-\frac{1}{n_{0}}\left\langle \int\mathrm{d}^3\mathbf{v}J_{0}e^{-iQ_{\kx}}\overline{J_{0} e^{iQ_{\kx}}} F_0 \right\rangle_\psi}, 
    \tag{\ref{eq:phi_infty}}
\end{equation}
where the only assumption is that 
\[
\left\langle \phi_{\infty}\int\mathrm{d}^3\mathbf{v}\overline{J_{0}e^{-iQ_{\kx}}}J_{0} e^{iQ_{\kx}} F_0 \right\rangle_\psi = \left\langle \int\mathrm{d}^3\mathbf{v}J_{0}e^{iQ_{\kx}}\overline{J_{0} e^{-iQ_{\kx}}} F_0 \right\rangle_\psi\langle\phi_{\infty}\rangle_\psi.
\]
It is useful here to apply the equivalence relation $\langle\int\mathrm{d}^3\mathbf{v}\dots\rangle_\psi = \langle\int\mathrm{d}^3\mathbf{v}\overline{\dots}\rangle_\psi$ \cite[equation~(D2)]{plunk2024residual}. This is the general expression used in the text.

\subsection{Moments of the Weakly Nonlinear Zonal Flow}
Here, we reinstate the $\veck$-indices, but continue to omit the species index. To find the moments in the main text, Eqs.~(\ref{eq:dn}) and (\ref{eq:dT}), the idealised form of the zonal distribution function in the weakly non-zonal regime, Eqs.~\eqref{eq:zonal} and \eqref{eq:nonzonal}, 
\begin{equation}
    \langle \df_{\kx}\rangle_\mathbf{R} = -\hat{\mathcal{N}} J_{0\ky}J_{0 \kx\ky}\left(\hat{v}^2-\frac{3}{2}\right) F_0,
\end{equation}
the key necessary integral can be written in the following form,
\begin{equation}
    I = \frac{1}{n_{0}}\int\mathrm{d}^3\mathbf{v}J_{0\kx}J_{0 \kx\ky}J_{0\ky}\left(\hat{v}^2-\frac{3}{2}\right) F_0. \label{eqn:integ_app_dn}
\end{equation}
It is natural to evaluate in the $v_\parallel$ and $v_\perp$ coordinates, as the arguments of the bessel functions read $J_{0\veck}=J_0(\sqrt{2b_{\veck}\xperp})$. Each of these integrals is separable.

Let us first proceed with the $\hat{v}_{\perp}$ integral. Integrals involving the product of three Bessel functions of the first kind can be expressed as infinite sums of modified Bessel functions \cite[equation~(3.8)]{glasser1994some}. Given that we shall eventually consider the small FLR limit, we shall expand Bessel functions according to $J_{0 \veck}\approx 1-b_{\veck}\xperp/2$ to order $O(b_{\veck})$  \citep{howes2006astrophysical}. To evaluate the perpendicular integrals we simply use
\begin{equation}
    \Sigma_n\stackrel{\cdot}{=}\int_0^\infty \hat{v}_{\perp}^{2n+1}e^{-\xperp}\mathrm{d}\hat{v}_{\perp}=\frac{n!}{2},
\end{equation}
so integral $I$ is given by,
\begin{equation}
\begin{split}
    I& \approx 2\left[-\Sigma_0+\left(1+\frac{b_{\kx}+b_{\kx \ky}+b_{\ky}}{2}\right)\Sigma_1-\frac{1}{2}(b_{\kx}+b_{\kx \ky}+b_{\ky})\Sigma_2\right]
    \\
    & =-\frac{b_{\kx}+b_{\kx \ky}+b_{\ky}}{2}.
\end{split}
\end{equation}
This is precisely the expression needed to find $\delta n$ in Eq.~(\ref{eq:dn}), which also gives us an expression for the potential associated to the perturbed distribution function, through quasineutrality. The expression for $\delta T$ is simpler to compute, as to leading order, the finite Larmor effects may be dropped, and thus all is left is Gaussian integrals in $v$.

\subsection{Residual of the Weakly Nonlinear Zonal Flow}
To compute the residual, we must now evaluate the integrals in equation~(\ref{eq:phi_infty}). These are significantly harder than those involved in the calculation of the moments in the preceding section, mainly due to the presence of the orbit width $Q_{\kx}$ and the bounce average. With this in mind, let us cautiously proceed, and start by expanding the integral in the denominator assuming small FLR and FOW effects. The denominator then takes the form,
\begin{equation}
    1-\frac{1}{n_{0}}\left\langle \int\mathrm{d}^3\mathbf{v}J_{0\kx}e^{-iQ_{\kx}}\overline{J_{0\kx} e^{iQ_{\kx}}} F_0 \right\rangle_\psi \approx \left\langle b_{\kx}\right\rangle_\psi-\underbrace{\frac{1}{n_{0}}\left\langle \int\mathrm{d}^3\mathbf{v}\left(\overline{Q_{\kx}}^2-\overline{Q_{\kx}^2}\right)F_0 \right\rangle_\psi}_{\stackrel{\cdot}{=}\mathcal{I}_Q}, \label{eqn:separate_flr_res_integral}
\end{equation}
where we used,
\begin{equation}
    \frac{1}{n_{0}}\left\langle\int \mathrm{d}^3\mathbf{v}J_{0\kx}^2 F_0 \right\rangle_\psi=\left\langle e^{-b_{\kx}}I_0(b_{\kx})\right\rangle_{\x} \approx 1-\langle b_{\kx}\rangle_\psi.
\end{equation}
To further proceed with the integral $\mathcal{I}_Q$, we must then devote some words to the form of $Q_{\kx}$ itself (full details may be found in \cite{Rodriguez_Plunk_2025}), and in particular its velocity space dependence. As it directly follows from its definition, $Q_{\kx}$ is linear in $v$ and the sign of $v_\parallel$. Otherwise, it is a function of both $\lambda$ and $\ell$ (field-line coordinate). Upon bounce average then, $\overline{Q_{\kx}}^2= v^2 f(\lambda) $ for some function $f$.

With this in mind, it is then convenient to rewrite $\mathcal{I}_Q$ as an integral over $v$ and $\lambda$,
\begin{equation}
    \mathcal{I}_Q=\frac{2}{\sqrt{\pi}}\left(\int\frac{\mathrm{d}\ell}{B}\right)^{-1}\int_0^\infty\mathrm{d}v v^4e^{-v^2}\int_0^{1/B_\mathrm{min}}\mathrm{d}\lambda\tau_b(\lambda)\frac{\overline{Q_{\kx}}^2-\overline{Q_{\kx}^2}}{v^2}, \label{eqn:I_Q}
\end{equation}
where the integrals over $\lambda$ and $v$ can be taken separately. We have here taken for simplicity there to be one single distinguishable well (like in a tokamak), but this can be easily relaxed by summing over wells and accordingly defining $B_\mathrm{min}$ for each of these. Thankfully, we do not need to evaluate this integral anew, as this is in fact the integral that \cite{RosenbluthHinton1998} solved to obtain their famous 1.6 numerical factor using a circular cross section axisymmetric geometry. In fact, $\mathcal{I}^\mathrm{RH}_Q=1.6q^2\left\langle b_{\kx}\right\rangle_\psi/ \sqrt{\Delta}$, where $\Delta$ is the mirror ratio.

Let us now look at the numerator of Eq.~(\ref{eq:phi_infty}), and proceed similarly with the expansion, keeping only terms at most quadratic in $Q_{\kx}$ or $k_\perp\rho_i$. Taking for simplicity $\hat{\mathcal{N}}$ to be constant along the fieldline\footnote{Of course, in actuallity, the distribution will have a ceratin fieldline dependence. However, we here look for what is the dominant effect, which remains in the velocity space.},

\begin{equation}
\begin{split}
    \frac{1}{n_{0}}\left\langle \int\mathrm{d}^3\mathbf{v}J_{0}e^{-iQ_{\kx}}\overline{e^{iQ_{\kx}}\frac{\langle\delta\!f_{\veck 0} \rangle_\mathbf{R}}{\hat{\mathcal{N}}}} \right\rangle_\psi \approx \underbrace{\frac{1}{n_{0}}\left\langle \int\mathrm{d}^3\mathbf{v}J_{0\kx}\overline{J_{0\ky}J_{0 \kx\ky}} f(v)F_0\right\rangle_\psi}_{\textcircled{1}} \\
    +  \underbrace{\frac{1}{n_{0}}\left\langle \int\mathrm{d}^3\mathbf{v}\left(\overline{Q_{\kx}}^2-\overline{Q_{\kx}^2}\right)f(v)F_0 \right\rangle_\psi}_{\textcircled{2}}.
\end{split} 
\end{equation}

The \textcircled{1} integral is precisely of the form used to compute $\delta n$, Eq.~(\ref{eqn:integ_app_dn}), if we overlook the bounce average. That is, we shall ignore the variation of the Larmor radius effect along the fieldline. In that case, we may use $\textcircled{1}\approx\langle\delta n/n\rangle_\psi=-\langle(b_{\kx}+b_{\ky} + b_{\kx \ky})/2\rangle$. The integral including the finite orbit width, \textcircled{2}, is harder. However, by writing it in the form of Eq.~(\ref{eqn:I_Q}), and noting that the only difference between the RH form of the integral and that needed here is the involvement of $f(v)$, we may simply relate,
\begin{equation}
    \textcircled{2}=\frac{\int_0^\infty x^4(x^2-3/2)e^{-x^2}\mathrm{d}x}{\int_0^\infty x^4e^{-x^2}\mathrm{d}x}\mathcal{I}_Q^\mathrm{RH}=\mathcal{I}_Q^\mathrm{RH}.
\end{equation}
This integral does not change at all! Therefore, we may write the residual for the weakly nonlinear zonal flow as,
\begin{equation}
    \phi_{\kx \infty}=-\frac{\langle(b_{\kx}+b_{\kx \ky}+b_{\ky})/2\rangle+\mathcal{I}_Q^\mathrm{RH}}{\langle b_{\kx}\rangle+\mathcal{I}_Q^\mathrm{RH}},
\end{equation}
which is the form used in the text in Eq.~\eqref{eq:residual_wstar} and in Figure~\ref{fig:residual_compare} for the analytic comparison. 

\section{Numerical Studies}\label{sec:numerical_studies}

This section summarises the numerical setups used to generate the figures presented in this paper and provides the information required to reproduce the results. The corresponding code and input files are available in the GitHub repositories \url{https://github.com/stellaGK/stella/tree/stella_weakly_nl} where the input files can be found in the sub-directory \texttt{INPUT\_FILES\_WEAKLYNL}.

\emph{Figure~\ref{fig:phi2_spectra}}: 
For this calculation, the nonlinear term was disabled for the non-zonal modes while remaining active for the zonal components. This behaviour is enabled by setting \texttt{weakly\_nonlinear = .true.} under the namelist \texttt{\&scale\_gyrokinetic\_terms}. The corresponding input file in the repository is \texttt{weakly\_nonlinear.in}.

\emph{Figure~\ref{fig:different_initial_conditions}}: The initialisation procedure allows restart files in which only the zonal modes are rescaled. This is controlled through the parameter \texttt{fac\_zonal}, located under the namelist \texttt{\&debug\_flags}, which can be varied between \texttt{0.0} and \texttt{1.0}. The corresponding input file is \texttt{scale\_zonal.in}.

\emph{Figure~\ref{fig:convolution}}: This figure was produced by varying the temperature gradient in an otherwise standard nonlinear simulation. The corresponding input file is \texttt{scan\_tprim.in}.

\emph{Figure~\ref{fig:residual_compare}}: This was produced by setting the RH initial condition to the analytic approximation predicted in equation Eq.~\eqref{eq:NL_term_residual}. This can be done by changing the toggle \texttt{weaknl\_rh} to \texttt{.True.} under the namelist \texttt{\&initialise\_distribution\_rh}. This input file is \texttt{RH\_init.in}.

\bibliography{jpp-instructions.bib}

@article{RosenbluthHinton1998,
    title   = {Poloidal flow driven by ion-temperature-gradient turbulence in tokamaks},
    author  = {Rosenbluth, M.N. and Hinton, F.L.},
    journal = {Phys. Rev. Lett.},
    volume  = {80},
    number  = {4},
    pages   = {724--727},
    year    = {1998}
}

@article{Hasegawawakatani,
    title = {Self-organization of electrostatic turbulence in a cylindrical plasma},
    author = {Hasegawa, A. and Wakatani, M.},
    journal = {Phys. Rev. Lett.},
    volume = {59},
    issue = {14},
    pages = {1581--1584},
    numpages = {0},
    year = {1987}
}

@article{Diamond_2005,
    year = {2005},
    month = {apr},
    publisher = {},
    volume = {47},
    number = {5},
    pages = {R35},
    author = {Diamond, P.H. and Itoh, S.-I. and Itoh, K. and Hahm, T.S.},
    title = {Zonal flows in plasma—a review},
    journal = {Plasma Physics and Controlled Fusion}
}

@article{glasser1994some,
  title={Some integrals involving Bessel functions},
  author={Glasser, M. Lawrence and Montaldi, E.},
  journal={Journal of Mathematical Analysis and Applications},
  volume={183},
  number={3},
  pages={577--590},
  year={1994},
  publisher={Elsevier}
}

@article{ItohItohDiamondetal,
    author  = {Itoh, K. and Itoh, S.-I. and Diamond, P.H. and Hahm, T.S.
               and Fujisawa, A. and Tynan, G.R. and Yagi, M. and Nagashima, Y.},
    title   = {Physics of zonal flows},
    journal = {Phys. Plasmas},
    volume  = {13},
    number  = {5},
    pages   = {055502},
    year    = {2006},
    doi     = {10.1063/1.2178779}
}

@article{friemanchen,
    author = {Frieman, E.A. and Chen, L.},
    title = {Variational method for electromagnetic waves in a magneto-plasma},
    journal = {Phys. Fluids},
    volume = {25},
    pages = {502},
    year = {1982}
}

@article{kruskalkulsrud,
    author = "Kruskal, M.D. and Kulsrud, R.M.",
    title = "Equilibrium of a Magnetically Confined Plasma in a Toroid",
    journal = "Phys. Fluids",
    year = "1958",
    volume = "1",
    number = "4",
    pages = "265-274"
}

@article{ChenLinWhite,
    author = {Chen, L. and Lin, Z. and White, R.},
    title = {Excitation of zonal flow by drift waves in toroidal plasmas},
    journal = {Phys. Plasmas},
    volume = {7},
    number = {8},
    pages = {3129-3132},
    year = {2000}
}

@article{Steve1991,
    author = {Cowley, S.C. and Kulsrud, R.M. and Sudan, R.},
    title = {Considerations of ion‐temperature‐gradient‐driven turbulence},
    journal = {Phys. Fluids B: Plasma Physics},
    volume = {3},
    number = {10},
    pages = {2767-2782},
    year = {1991},
    month = {10}
}

@article{Rodriguez_Plunk_2025, title={The zonal-flow residual does not tend to zero in the limit of small mirror ratio}, volume={91}, DOI={10.1017/S0022377825000066}, number={4}, journal={Journal of Plasma Physics}, author={Rodriguez, E. and Plunk, G.G.}, year={2025}, pages={E102}}

@article{BillGreg1993,
    author  = {Dorland, W. and Hammett, G.W.},
    title   = {Gyrofluid turbulence models with kinetic effects},
    journal = {Phys. Fluids B: Plasma Physics},
    volume  = {5},
    number  = {3},
    pages   = {812--835},
    year    = {1993}
}

@article{Rogers2000,
  title = {Generation and Stability of Zonal Flows in Ion-Temperature-Gradient Mode Turbulence},
  author = {Rogers, B.N. and Dorland, W. and Kotschenreuther, M.},
  journal = {Phys. Rev. Lett.},
  volume = {85},
  issue = {25},
  pages = {5336--5339},
  numpages = {0},
  year = {2000}
}

@unpublished{Richard2024,
    author = {Nies, R. and Parra, F.I. and Barnes, M. and Mandell, N. and Dorland, W.},
    title  = {Saturation of magnetized plasma turbulence by propagating zonal flows},
    note   = {Accepted for publication in Physical Review Research, 11 February 2026},
    year   = {2026}
}

@article{st-onge_2017, 
    title={On non-local energy transfer via zonal flow in the Dimits shift}, 
    volume={83},
    number={5}, 
    journal={Journal of Plasma Physics}, publisher={Cambridge University Press}, 
    author={St{-}{O}nge, D.A.},
    year={2017}
}

@article{ChenQiuZonca2024,
    author = {Chen, L. and Qiu, Z. and Zonca, F.},
    title = {On beat{-}driven and spontaneous excitations of zonal flows by drift waves},
    journal = {Phys. Plasmas},
    volume = {31},
    number = {4},
    pages = {040701},
    year = {2024}
}

@article{Zocco_Mishchenko_Koenies_Falessi_Zonca_2023, 
    title={Nonlinear drift-wave and energetic particle long-time behaviour in stellarators: solution of the kinetic problem}, 
    volume={89},  
    number={3}, 
    journal={Journal of Plasma Physics},
    author={Zocco, A. and Mishchenko, A. and K{\"o}nies, A. and Falessi, M. and Zonca, F.}, 
    year={2023}
}

@article{Barnes19,
	author = {{Barnes}, M. and {Parra}, F.I. and {Landreman}, M.},
	title = "{stella: An operator-split, implicit-explicit {\ensuremath{\delta}}f-gyrokinetic code for general magnetic field configurations}",
	journal = {Journal of Computational Physics},
	year = 2019,
	month = aug,
	volume = {391},
	pages = {365-380},
	doi = {10.1016/j.jcp.2019.01.025},
}

@article{zocco2018,
    title={Threshold for the destabilisation of the ion-temperature-gradient mode in magnetically confined toroidal plasmas},
    volume={84},
    OPTDOI={10.1017/S0022377817000988},
    number={1},
    journal={Journal of Plasma Physics},
    OPTpublisher={Cambridge University Press},
    author={Zocco, A. and Xanthopoulos, P. and Doerk, H. and Connor, J.W. and Helander, P.},
    year={2018},
    pages={715840101}
}

@article{Biglari89,
    author = {Biglari, H. and Diamond, P.H. and Rosenbluth, M.N.},
    title = {Toroidal ion-pressure-gradient-driven drift instabilities and transport revisited},
    journal = {Phys. of Fluids B},
    volume = {1},
    number = {1},
    pages = {109-118},
    year = {1989}
}

@book{gradshteyn2014table,
    title={Table of integrals, series, and products},
    author={Gradshteyn, I.S. and Ryzhik, I.M.},
    year={2014},
    publisher={Academic press}
}

@article{plunk2024residual,
  title={The residual flow in well-optimized stellarators},
  author={Plunk, G.G. and Helander, P.},
  journal={Journal of Plasma Physics},
  volume={90},
  number={2},
  pages={905900205},
  year={2024},
  publisher={Cambridge University Press}
}

@article{XiaoCatto2006,
  author  = {Xiao, Y. and Catto, P.J.},
  title   = {Short-wavelength effects on the collisionless neoclassical polarization and residual zonal flow level},
  journal = {Phys. Plasmas},
  volume  = {13},
  number  = {10},
  pages   = {102311},
  year    = {2006},
  doi     = {10.1063/1.2353814}
}

@article{SugamaWatanabe2006,
  author  = {Sugama, H. and Watanabe, T.-H.},
  title   = {Collisionless damping of zonal flows in helical systems},
  journal = {Phys. Plasmas},
  volume  = {13},
  number  = {1},
  pages   = {012501},
  year    = {2006},
  doi     = {10.1063/1.2164807}
}

@phdthesis{Nies2025Thesis,
  author       = {Nies, R.},
  title        = {Turbulence and flows in toroidal fusion plasmas},
  school       = {Princeton University},
  year         = {2025},
  url          = {https://dataspace.princeton.edu/handle/88435/dsp01p2677003v}
}

@book{kadomtsev2012reviews,
  title={Reviews of Plasma Physics},
  author={Kadomtsev, B.B. and Shafranov, V.D.},
  volume={21},
  year={2012},
  publisher={Springer Science \& Business Media}
}

@article{waltz2008numerical,
  title={Numerical experiments on the drift wave--zonal flow paradigm for nonlinear saturation},
  author={Waltz, R.E. and Holland, C.},
  journal={Phys. Plasmas},
  volume={15},
  number={12},
  year={2008},
  publisher={AIP Publishing}
}

@article{Winsor1968,
  author = {Winsor, N. and Johnson, J.L. and Dawson, J.M.},
  title = {Geodesic Acoustic Waves in Hydromagnetic Systems},
  journal = {Phys. Fluids},
  volume = {11},
  pages = {2448--2450},
  year = {1968},
  doi = {10.1063/1.1691790}
}

@article{Qiu2019,
  author = {Qiu, Z. and Chen, L. and Zonca, F.},
  title = {Kinetic theory of geodesic acoustic modes in toroidal plasmas: A brief review},
  journal = {Plasma Physics and Controlled Fusion},
  volume = {61},
  number = {1},
  pages = {014001},
  year = {2019},
  doi = {10.1088/1361-6587/aae72e}
}

@article{strintzi2007relation,
  title={On the relation between secondary and modulational instabilities},
  author={Strintzi, D. and Jenko, F.},
  journal={Phys. Plasmas},
  volume={14},
  number={4},
  year={2007},
  publisher={AIP Publishing}
}

@book{abramowitz1948handbook,
  title={Handbook of mathematical functions with formulas, graphs, and mathematical tables},
  author={Abramowitz, M. and Stegun, I.A.},
  volume={55},
  year={1948},
  publisher={US Government printing office}
}

@article{biglari1990influence,
  title={Influence of sheared poloidal rotation on edge turbulence},
  author={Biglari, H. and Diamond, P.H. and Terry, P.W.},
  journal={Phys. Fluids B: Plasma Physics},
  volume={2},
  number={1},
  pages={1--4},
  year={1990},
  publisher={American Institute of Physics}
}

@article{terry2000suppression,
  title={Suppression of turbulence and transport by sheared flow},
  author={Terry, P.W.},
  journal={Reviews of Modern Physics},
  volume={72},
  number={1},
  pages={109},
  year={2000},
  publisher={APS}
}

@article{lin1998turbulent,
  title={Turbulent transport reduction by zonal flows: Massively parallel simulations},
  author={Lin, Z. and Hahm, T.S. and Lee, W.W. and Tang, W.M. and White, R.B.},
  journal={Science},
  volume={281},
  number={5384},
  pages={1835--1837},
  year={1998},
  publisher={American Association for the Advancement of Science}
}

@article{tiwari2025zonal,
  title={Zonal flow suppression of turbulent transport in the optimized stellarators W7-X and QSTK},
  author={Tiwari, A. and Das, J. and Kumar, A.J. and Roberg-Clark, G. and Plunk, G.G. and Xanthopoulos, P. and Sharma, S. and Lin, Z. and Kuley, A.},
  journal={Plasma Physics and Controlled Fusion},
  volume={67},
  number={8},
  pages={085025},
  year={2025},
  publisher={IOP Publishing}
}

@article{monreal2016residual,
  title={Residual zonal flows in tokamaks and stellarators at arbitrary wavelengths},
  author={Monreal, P. and Calvo, I. and S{\'a}nchez, E. and Parra, F.I. and Bustos, A. and K{\"o}nies, A. and Kleiber, R. and G{\"o}rler, T.},
  journal={Plasma Physics and Controlled Fusion},
  volume={58},
  number={4},
  pages={045018},
  year={2016},
  publisher={IOP Publishing}
}

@article{xanthopoulos2011zonal,
  title={Zonal flow dynamics and control of turbulent transport in stellarators},
  author={Xanthopoulos, P. and Mischchenko, A. and Helander, P. and Sugama, H. and Watanabe, T.-H.},
  journal={Phys. Rev. Lett.},
  volume={107},
  number={24},
  pages={245002},
  year={2011},
  publisher={APS}
}

@techreport{watanabe2008regulation,
  title={Regulation of turbulent transport in neoclassically optimized helical configurations with radial electric fields},
  author={Watanabe, T.-H. and Sugama, H. and Ferrando-Margalet, S. and others},
  year={2008},
  institution={National Inst. for Fusion Science, Toki, Gifu (Japan)}
}

@article{mynick2010optimizing,
  title={Optimizing stellarators for turbulent transport},
  author={Mynick, H.E. and Pomphrey, N. and Xanthopoulos, P.},
  journal={Phys. Rev. Lett.},
  volume={105},
  number={9},
  pages={095004},
  year={2010},
  publisher={APS}
}

@article{goodman2024quasi,
  title={Quasi-isodynamic stellarators with low turbulence as fusion reactor candidates},
  author={Goodman, A.G. and Xanthopoulos, P. and Plunk, G.G. and Smith, H. and N{\"u}hrenberg, C. and Beidler, C.D. and Henneberg, S.A. and Roberg-Clark, G. and Drevlak, M. and Helander, P.},
  journal={PRX Energy},
  volume={3},
  number={2},
  pages={023010},
  year={2024},
  publisher={APS}
}

@article{smolyakov2000zonal,
  title={Zonal flow generation by parametric instability in magnetized plasmas and geostrophic fluids},
  author={Smolyakov, A.I. and Diamond, P.H. and Shevchenko, V.I.},
  journal={Phys. Plasmas},
  volume={7},
  number={5},
  pages={1349--1351},
  year={2000},
  publisher={American Institute of Physics}
}

@article{plunk2017nonlinear,
  title={Nonlinear growth of zonal flows by secondary instability in general magnetic geometry},
  author={Plunk, G.G. and Ba{\~n}{\'o}n Navarro, A.},
  journal={New Journal of Physics},
  volume={19},
  number={2},
  pages={025009},
  year={2017},
  publisher={IOP Publishing}
}

@article{qiu2016effects,
  title={Effects of energetic particles on zonal flow generation by toroidal Alfv{\'e}n eigenmode},
  author={Qiu, Z. and Chen, L. and Zonca, F.},
  journal={Phys. Plasmas},
  volume={23},
  number={9},
  year={2016},
  publisher={AIP Publishing}
}

@article{brizard2007foundations,
  title={Foundations of nonlinear gyrokinetic theory},
  author={Brizard, A.J. and Hahm, T.S.},
  journal={Reviews of modern physics},
  volume={79},
  number={2},
  pages={421--468},
  year={2007},
  publisher={APS}
}

@article{abel2013multiscale,
  title={Multiscale gyrokinetics for rotating tokamak plasmas: fluctuations, transport and energy flows},
  author={Abel, I.G. and Plunk, G.G. and Wang, E. and Barnes, M. and Cowley, S.C. and Dorland, W. and Schekochihin, A.A.},
  journal={Reports on Progress in Physics},
  volume={76},
  number={11},
  pages={116201},
  year={2013},
  publisher={IOP Publishing}
}

@article{hammett1993developments,
  title={Developments in the gyrofluid approach to tokamak turbulence simulations},
  author={Hammett, G.W. and Beer, M.A. and Dorland, W. and Cowley, S.C. and Smith, S.A.},
  journal={Plasma physics and controlled fusion},
  volume={35},
  number={8},
  pages={973--985},
  year={1993}
}

@article{todo2010nonlinear,
  title={Nonlinear magnetohydrodynamic effects on Alfv{\'e}n eigenmode evolution and zonal flow generation},
  author={Todo, Y. and Berk, H.L. and Breizman, B.N.},
  journal={Nuclear Fusion},
  volume={50},
  number={8},
  pages={084016},
  year={2010}
}

@article{nies2026theory,
  title={Theory of zonal flow growth and propagation in toroidal geometry},
  author={Nies, R. and Parra, F.I.},
  journal={Plasma Physics and Controlled Fusion},
  volume={68},
  number={4},
  pages={045028},
  year={2026},
  publisher={IOP Publishing}
}

@article{ivanov2023analytical,
  title={An analytical form of the dispersion function for local linear gyrokinetics in a curved magnetic field},
  author={Ivanov, P.G. and Adkins, T.},
  journal={Journal of Plasma Physics},
  volume={89},
  number={2},
  pages={905890213},
  year={2023},
  publisher={Cambridge University Press}
}

@article{hall1975three,
  title={Three-dimensional equilibrium of the anisotropic, finite-pressure guiding-center plasma: Theory of the magnetic plasma},
  author={Hall, L.S. and McNamara, B.},
  journal={The Phys. Fluids},
  volume={18},
  number={5},
  pages={552--565},
  year={1975},
  publisher={AIP Publishing}
}

@article{cary1997omnigenity,
  title={Omnigenity and quasihelicity in helical plasma confinement systems},
  author={Cary, J.R. and Shasharina, S.G.},
  journal={Phys. Plasmas},
  volume={4},
  number={9},
  pages={3323--3333},
  year={1997},
  publisher={American Institute of Physics}
}

@article{catto1977linearized,
  title={Linearized gyro-kinetic equation with collisions},
  author={Catto, P.J. and Tsang, K.T.},
  journal={Phys. Fluids},
  volume={20},
  number={3},
  pages={396--401},
  year={1977}
}

@article{howes2006astrophysical,
  title={Astrophysical gyrokinetics: basic equations and linear theory},
  author={Howes, G.G. and Cowley, S.C. and Dorland, W. and Hammett, G.W. and Quataert, E. and Schekochihin, A.A.},
  journal={The Astrophysical Journal},
  volume={651},
  number={1},
  pages={590--614},
  year={2006}
}

@article{guillon2025phase,
  title={Phase transition from turbulence to zonal flows in the Hasegawa--Wakatani system},
  author={Guillon, P.L. and G{\"u}rcan, {\"O}.D.},
  journal={Phys. Plasmas},
  volume={32},
  number={1},
  year={2025},
  publisher={AIP Publishing}
}

@article{mishchenko2008collisionless,
  title={Collisionless dynamics of zonal flows in stellarator geometry},
  author={Mishchenko, A. and Helander, P. and K{\"o}nies, A.},
  journal={Phys. Plasmas},
  volume={15},
  number={7},
  year={2008},
  publisher={AIP Publishing}
}

@article{helander2011oscillations,
  title={Oscillations of zonal flows in stellarators},
  author={Helander, P. and Mishchenko, A. and Kleiber, R. and Xanthopoulos, P.},
  journal={Plasma Physics and Controlled Fusion},
  volume={53},
  number={5},
  pages={054006},
  year={2011}
}
\bibliographystyle{jpp}

\end{document}